\documentclass[showpacs,preprintnumbers]{revtex4}
\usepackage{amsfonts}
\usepackage{amssymb}
\usepackage{graphicx}
\usepackage{color}
\usepackage{amsmath}

\begin{document}

\title{Reanalysis of $^{13}$N($p,\gamma $)$^{14}$O reaction and its role in
stellar CNO cycle}
\author{S. B. Dubovichenko,$^{1,2}$ R. Ya. Kezerashvili,$^{3,4}$ N. A.
Burkova,$^{2}$ A. V. Dzhazairov-Kakhramanov$^{1}$, and B. Beisenov$^{1,2}$}
\affiliation{$^{1}$Fesenkov Astrophysical Institute \textquotedblleft
NCSRT\textquotedblright\ ASA MDASI RK, 050020, Almaty, Kazakhstan\\
$^{2}$al-Farabi Kazakh National University, 050040, Almaty, Kazakhstan\\
$^{3}$New York City College of Technology, City University of New York,
Brooklyn, NY 11201, USA\\
$^{4}$Graduate School and University Center, City University of New York,
New York 10016, USA}
\date{\today}

\begin{abstract}
Within the framework of the modified potential cluster model with forbidden
states, the $^{13}$N($p,\gamma $)$^{14}$O reaction rate and the
astrophysical $S$-factor are considered. It is shown that the first $p^{13}$%
N resonance determines the $S$-factor and contributions of the $M1$ and $E2$
transitions are negligible at energies $E<1$ MeV, but are significant at
high energies. The $S$-factor strongly depends on the $^{3}S_{1}$ resonance
parameters. The influence of the width of \ the $^{3}S_{1}$ resonance on $S$%
-factor is demonstrated. The reaction rate is calculated and an analytical
approximation for the reaction rate is proposed. A comparison of our
calculation with existing data is addressed. Results of our calculations for
the $^{13}$N($p,\gamma )^{14}$O reaction rate provide the contribution to
the steadily improving reaction rate database libraries. Our calculations of
the $^{13}$N($p,\gamma )^{14}$O reaction rate along with results for the
rates of $^{14}$N($p,\gamma )^{15}$O and $^{12}$C$(p,\gamma )^{13}$N
processes provide the temperature range $0.13<T_{9}<0.97$ for the conversion
of CNO cycle to the HCNO cycle. Our results demonstrate that at early stages
of a nova explosion at temperatures about $0.1$ $T_{9}$ and at late stages
of evolution of supermassive stars at temperatures about $1.0$ $T_{9}$ the
ignition of the HCNO cycle could occur at much lower densities of a stellar
medium.
\end{abstract}

\maketitle


\section{Introduction}

Radiative capture reactions play an important role in astrophysics.
Light elements are either created during the big bang or during fusion
reactions in stars. In the latter case, they are the result of hydrogen
burning which is characterized by two major reaction sequences: i. the $pp$
chain; ii. the carbon-nitrogen-oxygen (CNO) cycles ~\cite{1}. The CNO cycle
is considered as a catalytic process that requires the presence of some
initial carbon, nitrogen, and oxygen abundance in the stellar material.
Radiative capture reactions, namely those in which an atomic nucleus fuses
with one proton or neutron and produces a nucleus with the emission of
electromagnetic radiation, or with $\alpha -$particle emission, have the
greatest importance in nuclear astrophysics \cite{Wiescher2012,Brune2015}.
In particular, competing $(p,\gamma )$ and $(p,\alpha )$ reactions are
branching points in the CNO cycling process \cite{1}. However, the
strong-interaction $(p,\alpha )$ branch is substantially stronger than the
electromagnetic $(p,\gamma )$ branch, but, in some cases, the latter one can
be comparable with the $(p,\alpha ),$ which alters the reaction flow
substantially in certain astrophysical temperature regimes \cite%
{Wiescher1980}. The proton induced radiative capture reactions $(p,\gamma )$
occur in many stellar environments, for example, in novae and $X-$ray
bursts. Especially in stellar environments due to the high temperatures and
short reaction times $(p,\gamma )$ reactions involving short-lived nuclei
play an important role for energy generation and nucleosynthesis. It takes
the high-density environment of stars to generate nuclei with masses $%
A\geqslant $12. The reactions of protons' radiative capture are widely
discussed in the literature (see reviews \cite{Wiescher2012,Brune2015,2} and
references herein). It is done primarily due to the fact that the carbon
component burns out in a series of processes known as hot CNO cycle
(HCNO-I), which occurs at temperatures starting from 0.2 $T_{9}$ \cite{1}.
The synthesized isotope $^{14}$O is considered as a waiting point, which is
overcome by a chain of reactions, starting with $^{14}$O$\left( \alpha
,p\right) ^{17}$F when temperature is above 0.4 $T_{9}$. The review \cite{1}
presents comprehensive and consistent illustrations of CNO and HCNO-I
cycle chains, as well as the evolution of the CNO isotope abundance with time for
different density and temperature conditions, the calculations of which are
directly based on the reaction rates.

The pioneering measurement with a rare-isotope beam was the first direct
determination of the $^{13}$N($p,\gamma $)$^{14}$O reaction cross section
using a radioactive $^{13}$N beam \cite{Decrock1991,22,Decrock1993}. In the
reaction $^{13}$N($p,\gamma $)$^{14}$O the $s-$wave capture on the broad 1$%
^{-}$ resonance dominates the reaction rate and over three decades many
efforts have been made to determine the parameters for resonance using
different experimental approaches: transfer reactions \cite%
{Chunpp1985,Fernandez1985,Smith,22}, Coulomb dissociation of high energy $%
^{14}$O beam in the field of a heavy nucleus \cite%
{Motobayashi1991,Kiener1993,Bauer1994}, a rare-isotope beam \cite%
{Decrock1991,22,Decrock1993}, using the unstable ion beam by indirect
measurements\ \cite{16,17}, and, most recently, via neutron-knockout
reactions with a fast $^{15}$O beam \cite{15}. Ref. \cite{2} provides an
overview of current experimental projects specializing in the synthesis of
radioactive isotope beams and experiments on astrophysical applications.
However, there is no experimental data today suitable for comparison with
theoretical calculations of cross sections or astrophysical $S$-factors. In
this case, apparently, it is possible to synthesize $^{13}$N isotope beams,
given that its lifetime of 9.965 min is comparable with the neutron
lifetime. At the same time, direct measurements of the $^{14}$O($\alpha ,p$)$%
^{17}$F reaction are carried out, although the $\beta ^{+}$ decay of isotope
$^{14}$O is 70.598 s.
Nevertheless, in the future we can expect new data for cross sections of
the process $^{13}$N($p,\gamma $)$^{14}$O \cite{2}.

The results of the studies \cite%
{27,26,Funck1987,Decrock1991,Decrock1993,23,16,17,18,25,Huang2010} on
astrophysical $S$-factor and $^{13}$N($p,\gamma $)$^{14}$O reaction rate are
included in the NACRE (Nuclear Astrophysics Compilation of REactions)
database \cite{28} and in the new compilation, referred to as NACRE II \cite%
{24}. These databases form the basis for macroscopic astrophysical
calculations. The key generalizing element of all calculations is the first $%
^{3}S_{1}$ resonance in the $p^{13}$N scattering channel and all
calculations are based on the energy and the width of this resonance. In the
above mentioned works, experimental data on these characteristics are taken
from Ajzenberg's 1991 compilation \cite{12}. At present, new data are
available on the spectra of $^{14}$O nucleus \cite{14}. Therefore, it is
relevant to consider these data for analysis of the $^{13}$N($p,\gamma $)$%
^{14}$O reaction. Moreover, another incentive for these calculations are the
data from the latest experimental research \cite{15} that will be also
brought to our discussion.

Theoretical calculations of a reaction rate rely on the reaction cross
section, which is determined by the nuclear structure of the nuclei
involved, the reaction mechanism, and the associated interaction forces. The
cross section can be calculated in the framework of \textit{ab initio}
models, where it is determined using the wave functions (WFs) of the system,
but subject to uncertainties associated with the theoretical model and the
quality of the optical potential. Most notable are cluster model approaches,
where nucleons are grouped in clusters of particles, which is a
configuration that might, in particular, enhance the reaction rates and that also 
rely on the quality of the optical potential \cite{4,Wiescher2017,3}.
Calculations of the rate for the $^{13}$N($p,\gamma $)$^{14}$O reaction and
the astrophysical $S$-factor were performed within potential models using a
shell-model, cluster model and $R$-matrix approaches \cite%
{27,26,Funck1987,18,25}. There are significant differences between the
various calculations of the $^{13}$N($p,\gamma $)$^{14}$O reaction as well
as in the light of a new experimental study \cite{15}, an independent and
well established approach is greatly needed to analyze this process.
Continuing our studies of the processes of radiative capture on light atomic
nuclei (see Refs. \cite{4,3,NucPhys2019,Dubovichenko2020} for concise
summaries), we consider the reaction of $p+^{13}$N$\rightarrow $ $^{14}$O$%
+\gamma $ at astrophysical energies. This process is clearly not included in
the thermonuclear standard CNO cycle, but it makes a certain contribution to
accumulation processes of a stable $^{14}$N nucleus, which is further
involved in other reactions of this cycle \cite{5} and belongs to the hot CNO
cycle \cite{1}.

The goal of this study is twofold: i. to calculate the cross section of the $%
^{13}$N($p,\gamma $)$^{14}$O reaction at the energies of astrophysical
interest and the reaction rate as a function of temperature for the analyses
of the influence of the first $p^{13}$N resonance width on the astrophysical
$S$-factor; ii. to analyze and determine a temperature range for the
conversion of the CNO cycle to the HCNO cycle.

The article is organized as follows. In Sec. II the potential cluster model
with the classification of orbital states and methods of calculations are
described. Classification and structure of states are introduced and
analyzed in Sec. III, while in Sec. IV the potentials for the $p^{13}$N
interaction are presented. Astrophysical $S$-factor of the proton radiative
capture on $^{13}$N and the $^{13}$N($p,\gamma $)$^{14}$O reaction rate are
given in Sec. V. The role of the $^{13}$N($p,\gamma $)$^{14}$O reaction in
the conversion from the CNO to the hot CNO cycle is discussed in Sec. VI.
Conclusions follow in Sec. VII.

\section{Theoretical model and formalism}

\bigskip To carry out calculations of astrophysical $S$-factors for various
reactions, we usually use the modified potential cluster model (MPCM) of
light atomic nuclei \cite{4,3,NucPhys2015,NucPhys2019,Dubovichenko2020} with
the classification of orbital states according to Young diagrams \cite{6,7}.
The model provides relatively many simple possibilities for performing
calculations of various astrophysical characteristics. For example, one can
calculate the astrophysical $S$-factor of radiative capture for
electromagnetic transitions from scattering states of clusters to bound
states (BS) of light atomic nuclei in cluster channels \cite{3,4}. The
choice of this model is due to the fact that in many atomic nuclei the
probability of cluster formation and the degree of their separation are
relatively high. This is confirmed by numerous experimental data and various
theoretical calculations obtained in various works over the past few decades
\cite{7}.

Thermonuclear rates are defined by reaction cross sections which can be
obtained using a theoretical model. In the present study of the $^{13}$N($%
p,\gamma $)$^{14}$O reaction we use the modified potential cluster model,
where a proton interacts with a system of nucleons which are grouped into
cluster $^{13}$N. States of the $p-^{13}$N system are defined by the
classification according to Young diagrams. Relative motion WFs are
determined by solving the Schr\"{o}dinger equation \cite%
{4,3,NucPhys2015,NucPhys2019}. The entry channel presents the proton $p$($%
\frac{1}{2}^{+}$) ($J^{\pi }$ is the total momentum and parity) and $^{13}$N(%
$\frac{1}{2}^{-}$) nucleus. For description of the final state we assume
that $^{14}$O nucleus consist of the same particles as in the initial
channel, but in the bound state.

In the microscopic formalism widely known as the resonating-group method
\cite{Wildermuth,RGM}, the wave function WFs of the $p^{13}$N system has the
form of an antisymmetrized product of internal cluster wave functions and a
WF of their relative motion:
\begin{equation}
\Psi =\hat{A}[\psi _{p}(\mathbf{r}_{1})\psi _{^{13}\text{N}}(\mathbf{r}%
_{2})\chi (\mathbf{r}_{1}-\mathbf{r}_{2})].  \label{fanction}
\end{equation}%
In Eq. (\ref{fanction}) $\hat{A}$ is the antisymmetrization operator, $\psi
_{p}(\mathbf{r}_{1})$ and $\psi _{^{13}\text{N}}(\mathbf{r}_{2})$ are the
wave functions of the proton and $^{13}$N nucleus, respectively, $\mathbf{r}%
_{1}$ and $\mathbf{r}_{2}$ are the radius vectors of their center of mass, $%
\chi (\mathbf{r})$ is the WF of their relative motion, while $\mathbf{r}=%
\mathbf{r}_{1}-\mathbf{r}_{2}.$

According to \cite{6,Wildermuth} the WF of $^{13}$N is antisymmetrized.
Thus, only exchange transpositions between nucleons of the $^{13}$N nucleus
and proton must be taken into account, which leads to the modification of the
function. By contrast, in our approach this method of antisymmetrization
consists in the effective accounting of the Pauli principle by using the
deep attractive potentials with the forbidden states (FS). Mathematically
this realization is based on the classification of orbital states according to the
Young diagrams \cite{6,7}. Exclusion of FSs from spectra leads to the
correct node behavior of the function in the internal range, both for a
bound state and for a continuous spectrum that, in its turn, reflects on the
asymptotics of these functions.

To build interaction potentials between the proton and $^{13}$N for
scattering states in the MPCM, results of phase shift analysis of
experimental data of differential cross sections for an elastic scattering
of corresponding particles are generally used. The other way to build the
potentials is to use spectra of the resulting nucleus $^{14}$O \cite{3,4}.
Moreover, the multiparticle nature of the problem is taken into account by
dividing single-particle levels of such a potential into allowed and
forbidden by Pauli principle \cite{6,7} states. The concept of
Pauli-forbidden states allows one to consider the multi-body character of
the problem in terms of two-body interaction potential between clusters.
Potentials for bound states (BS) of $p$ and $^{13}$N particles are built
primarily based on the requirement to describe the main characteristics of
the $^{14}$O nucleus. For example, this is a requirement to reproduce the
binding energy of $^{14}$O in a corresponding $p$$^{13}$N cluster channel
and a description of the other static nuclear characteristics, such as a
charge radius and asymptotic constant (AC), with the same potential \cite%
{NucPhys2015}. The functions of the initial $p$$^{13}$N and final $^{14}$O
states are characterized by specific quantum numbers, including the Young
diagrams {$f$}, which determine the permutation symmetry of the orbital part
of relative motion WFs of these states. Thus, the problem can be reduced to
two parts:

i. a construction of $p$$^{13}$N interaction potentials with the FS for each
partial wave, i.e., for the given orbital angular momentum $L$, which also
includes a point-like Coulomb term;

ii. the numerical solution of the radial Schr\"{o}dinger equation for these
potentials to find the corresponding WFs of the relative motion.

Further, following Refs. \cite{3,4,NucPhys2015,NucPhys2019,Dubovichenko2020}%
, we use well-known expressions for total cross sections and matrix elements
of multipole transition operators with the initial and final channel spins $%
S_{i}=S_{f}=S$

\begin{equation}
\sigma _{c}(NJ,J_{f})=\frac{8\pi Ke^{2}}{\hbar ^{2}k^{3}}\frac{\mu }{%
(2S_{1}+1)(2S_{2}+1)}\frac{J+1}{J\left[ (2J+1)!!\right] ^{2}}%
A_{J}^{2}(NJ,K)\sum_{L_{i}J_{i}}P_{J}^{2}\left( NJ,J_{f},J_{i}\right)
I_{J}^{2}(J_{f},J_{i}),  \label{Crossection}
\end{equation}%
where the notation $NJ$ corresponds to $EJ$ for the electric and $MJ$ for
the magnetic transitions, respectively. The matrix elements of the $EJ$
transitions have a form
\begin{equation}
P_{J}^{2}\left( EJ,J_{f},J_{i}\right)
=(2J+1)(2L_{i}+1)(2J_{i}+1)(2J_{f}+1)\left( L_{i}0J0\right\vert
L_{f}0)^{2}\left\{
\begin{array}{c}
L_{i}SJ_{i} \\
J_{f}JL_{f}%
\end{array}%
\right\} ^{2}  \label{Matelement}
\end{equation}%
and

\begin{equation}
A_{J}(EJ,K)=K^{J}\mu ^{J}\left( \frac{Z_{1}}{m_{1}^{J}}+(-1)^{J}\frac{Z_{2}}{%
m_{2}^{J}}\right) ,\text{ \ \ \ }  \label{Parameter1}
\end{equation}%
\begin{equation}
I_{J}(J_{f},J_{i})=\left\langle \chi _{f}\right\vert r^{J}\left\vert \chi
_{i}\right\rangle .  \label{Overlapintegr}
\end{equation}%
In Eqs. (\ref{Crossection}) - (\ref{Overlapintegr}) $e$ is the elementary
charge, $K=\frac{E_{\gamma }}{\hbar c}$ is the wave number of the emitted
photon with energy $E_{\gamma }$, $k$ is the wave number of particles in the
initial channel, $m_{1},$ $m_{2},$ $Z_{1},$ $Z_{2}$ and $\mu $ are masses,
charges of colliding nuclei and their reduced mass, respectively, in the
initial channel, $S_{i},$ $S_{f},$ $L_{i},$ $L_{f},$ $J_{i},$ $J_{f}$ are
the total spins, orbital momenta, total momenta of particles in the initial (%
$i$) and final ($f$) channels, respectively, while $\left(
L_{i}0J0\right\vert L_{f}0)$ are the Clebsch -- Gordan coefficients and $%
\left\{
\begin{array}{c}
... \\
...%
\end{array}%
\right\} $ are the $6j-$symbols. The integral $I_{J}(J_{f},J_{i})$ is
defined by using WFs of relative motion of particles in the initial $\chi
_{i}(r)$ and final $\chi _{f}(r)$ states, which depend on an intercluster
distance $r$.

In the general form for $MJ$ transitions for arbitrary rank $J$, the matrix
element in Eq. (\ref{Crossection}) can be written using the $9j-$symbols as

\begin{eqnarray}
P_{J}^{2}\left( MJ,J_{f},J_{i}\right)
&=&S(S+1)(2S+1)(2J_{i}+1)(2L_{i}+1)(2J-1)(2J+1)(2J_{f}+1)  \notag \\
&&\times \left( L_{i}0J-10\right\vert L_{f}0)^{2}\left\{
\begin{array}{c}
L_{i}J-1L_{f} \\
S\text{ \ \ }1\text{ \ \ }S \\
J_{i}\text{ \ }J\text{ \ }J_{f}%
\end{array}%
\right\} ^{2},  \label{Matelement2}
\end{eqnarray}

\begin{equation}
A_{J}(MJ,K)=\frac{\hbar K}{m_{0}c}K^{J-1}\sqrt{J(2J+1)}\left[ \mu _{1}\left(
\frac{m_{2}}{m}\right) ^{J}+(-1)^{J}\mu _{2}\left( \frac{m_{1}}{m}\right)
^{J}\right] ,\text{ }
\end{equation}

\begin{equation}
I_{J}(J_{f},J_{i})=\left\langle \chi _{f}\right\vert r^{J-1}\left\vert \chi
_{i}\right\rangle ,  \label{Overlapintegr2}
\end{equation}%
where $m$ is a mass of a nucleus in the final channel, $\mu _{1}$ and $\mu
_{2}$ are magnetic momenta of the clusters, and the remaining notations are
the same as in Eqs. (\ref{Crossection}) - (\ref{Parameter1}).

Thus, to find the cross section of the $^{13}$N($p,\gamma $)$^{14}$O
reaction one should calculate the expressions (\ref{Overlapintegr}) and (\ref%
{Overlapintegr2}) for $EJ$ and $MJ$ transitions, respectively. The latter
requires finding the radial WFs $\chi _{i}$ and $\chi _{f}$ of relative
motion of particles in the initial and final states.

\section{\protect\bigskip Classification and structure of states}

Let us now consider a classification of $p^{13}$N system orbital states
according to the Young diagram. It was previously shown that the ground
bound state (GS) of $^{13}$N and $^{13}$C nuclei corresponds to the Young
orbital diagram \{4441\} \cite{6,10}. Recall that possible Young's orbital
diagrams in the system of $N=n_{1}+n_{2}$ particles can be defined as a
direct external product of the orbital diagrams of each subsystem \cite%
{11,111K}, which for the $p^{13}$N system within 1$p$ shell gives $%
\{1\}\times \{4441\}\rightarrow \{5441\}+\{4442\}$. The first of the
obtained diagrams is compatible with orbital momentum $L=1,3$ and is
forbidden for the $s$-shell, since there cannot be five nucleons in the $s$%
-shell, while the second diagram is allowed and compatible with the orbital
momenta zero and two \cite{11,111K}. Thus, the potential of the $^{3}S_{1}$ {%
(here and below we use notations $^{2S+1}L_{J}$ for resonances) }wave has
only the allowed state, but the $P$ and $F$ waves have both forbidden and
allowed states \cite{12}. However, since we do not have complete tables of
the products of Young diagrams for a system with a number of particles
greater than eight \cite{13}, which we used earlier for such calculations
\cite{3,4}, the result obtained above should be considered only as a
qualitative estimate of possible orbital symmetries in the ground state of $%
^{14}$O nucleus for the $p^{13}$N channel.

We now consider the basic characteristics of $^{14}$O nucleus, which has in
the GS $J^{\pi }=0^{+}$ the energy 4.628 MeV \cite{12}. Since for the $^{13}$%
N nucleus $J^{\pi }=1/2^{-}$ \cite{12}, the GS of $^{14}$O in the $p^{13}$N
channel can be associated with the $^{3}P_{0}$ state. Below this threshold,
there are no bound excited states (ES) \cite{12}. Above the threshold, there
are the following resonance states (RS):

\bigskip 1. For the first resonance, which plays the most important role in
determining the magnitude of the astrophysical $S$-factor, new data \cite%
{15} lead to an excitation energy of 5.164(12) MeV (here and below \ numbers
in parentheses are uncertainties), which corresponds to the energy $%
E_{res}=536(12)$ keV relative to the threshold in the center-of-mass (c.m.),
the width $\Gamma _{res}=38(2)$ keV, and momentum $J^{\pi }=1^{-}$.
Previously in Ref. \cite{14} it was reported for this level the excitation
energy of 5.156(2) MeV, i.e\textit{.} $E_{res}=0.528(2)$ MeV and the width $%
\Gamma _{res}=37.3(9)$ keV. In an earlier work \cite{12}, for this resonance
the excitation energy 5.173(10) MeV, i.e. $E_{res}=545(10)$ keV and the
width $\Gamma _{res}=38.1(1.8)$ keV were reported. In fact, these three
results lead to the same 38(2) keV widths. However, the resonance energies
do not overlap within the experimental errors and can be in the range of $%
E_{res}=524-555$ keV. This resonance can be matched to the $^{3}S_{1}$
state, and $E1$ transition $^{3}S_{1}\rightarrow $ $^{3}P_{0}$ is possible.
It is clear that it cannot be $^{3}D_{1}$ because this needs protons in the 1%
$d_{3/2}$ shell (in the framework of a shell-model scheme), which is much
higher in energy and likely irrelevant for this state. In this paper, we
consider the $E1$ transition $^{3}S_{1}\rightarrow $ $^{3}P_{0}$.

All other resonances, as can be seen below, do not make a significant
contribution to the $S$-factor at low energies, and their energies, as
follows from Refs. \cite{15} and \cite{14}, practically overlap. Therefore
we use the data \cite{14}, but for a comparison we also give the energies
and widths obtained in Ref. \cite{15}.

\bigskip 2. At an excitation energy of 5.710(20) MeV or 1.082(20) MeV
relative to the channel's threshold in the c.m., there is a state $J^{\pi
}=0^{-}$ with a width of 400(45) keV \cite{14}, which can be associated with
a $^{1}S_{0}$ wave. However, in this case, the transition to the GS is
impossible, because it refers to a triplet state. Let us mention that the
classification of allowed transitions is defined by the algebra of geometric
addition of angular momenta, represented by the Clebsch-Gordan coefficients,
6${j}${\ and 9}${j-}$ symbols \cite{Varshalovich,Tkach2019}. Besides, $EJ$
and $MJ$ transitions change parity of the initial and final states according
to (-1)$^{j}$ and (-1)$^{j+1}$, respectively. So, for example, $%
^{1}S_{0}\rightarrow $ $^{2}P_{0}$ transition is not allowed because there
is no $E$ or $M$ transition connecting 0$^{-}$ and 0$^{+}$ states that is
seen from Eqs. (\ref{Matelement}) and (\ref{Matelement2}).

\bigskip 3. At an excitation energy of 5.920(10) MeV, i.e. $E_{res}=1.29(10)$
MeV, there is a state $J^{\pi }=1^{+}$ with a width $\Gamma _{res}<12$ keV
\cite{15}, which can be matched to a $^{3}P_{0}$ wave. In Ref. \cite{15} the
energy 5.931(10) MeV and the width less than 12 keV were reported. From this
wave, magnetic transitions to the GS are impossible.

\bigskip 4. At an excitation energy of 6.284(9) MeV [$E_{res}=1.656(9)$ MeV
in the c.m.], there is a state $J^{\pi }=3^{-}$ with the width $\Gamma
_{res}=$ $25(3)$ keV \cite{14}, while in Ref. \cite{15} the energy 6.285(12)
MeV and the width 37.7(17) keV are obtained. This state can be matched to a $%
^{3}D_{3}$ wave. From this wave, only the $E3$ transition is possible, which
is omitted in our consideration, because of its smallness.

\bigskip 5. At an excitation energy of 6.609(10) MeV [$E_{res}=1.981(10)$
MeV], there is a state $J^{\pi }=2^{+}$ with a width $\Gamma _{res}<5$ keV
\cite{14}, which can only be associated with a $^{3}P_{2}$ or $^{3}F_{2}$
waves. In Ref. \cite{15} the energy 6.585(11) MeV and the width less than 25
keV is reported. For $^{3}F_{2}$ wave the $E2$ transition is possible and we
evaluate its effect.

\bigskip 6. At an excitation energy of 6.767(11) MeV [$E_{res}=2.139(11)$
MeV], there is a state $J^{\pi }=2^{-}$ with the width $\Gamma _{res}<90(5)$
keV \cite{14}. Based on the results \cite{15}, the energy is 6.764(10) MeV
and the width is 96(5) keV. This state can be associated with a $^{3}D_{2}$
wave. From this wave, only $M2$ transition to the GS is possible. This
transition is omitted, because we restrict ourselves with the consideration
of the $M1$ transition only.

\bigskip 7. At an excitation energy of 7.768(10) MeV [3.140(10) MeV in the
c.m.] for the state $J^{\pi }=2^{+}$ the width of 68(6) keV was observed in
Ref. \cite{15}, while Ref. \cite{14} reported 7.745(19) MeV [$%
E_{res}=3.117(19)$] and 62(10) keV for the energy and the width,
respectively. This resonance state can be associated with the $^{3}P_{2}$ or
$^{3}F_{2}$ waves. From the $^{3}F_{2}$ wave the $E$2 transition to the GS
is possible and we evaluate its effect.

\bigskip 8. Recently, in Ref. \cite{15} at the excitation energy of \
9.755(10) MeV or 5.123(11) MeV relative to the threshold of the $p^{13}$N
channel, a state $J^{\pi }=2^{+}$ with the width $\Gamma _{res}=$229(51) keV
was observed. While the excitation energy is in good agreement with the
results from Ref. \cite{14}, 9.751(11) MeV, the width of the resonance is
almost twice bigger. Moreover, a momentum $J^{\pi }=2^{+}$ of this state was
in question in \cite{14}, but in the recent work \cite{15} it was finally
determined. This state can also be associated with $^{3}P_{2}$ or $^{3}F_{2}$
waves. From the $^{3}F_{2}$ wave the $E$2 transition to the GS is also
possible, and we will consider its effect.



As a result of the analysis of the above mentioned resonances,
it turns out that, first of all, it is necessary to consider the $E1$
transition from the first resonance at $E_{res}=$536(12) keV with $J^{\pi
}=1^{-}$ and the width $\Gamma _{res}=38(2)$ keV \cite{15}. In addition, we
consider two other values for the energy of this resonance $E_{res}=528(2)$
keV with the width $\Gamma _{res}=37.3(9)$ keV \cite{14} and $%
E_{res}=545(10) $ keV with the width $\Gamma _{res}=38.1(1.8)$ \cite{12}. In
addition to the $E1$ transition, there are three $E2$ transitions for $%
J^{\pi }=2^{+}$, $E_{res}=$1.981(10) MeV, $\Gamma _{res}=5$ keV, $J^{\pi
}=2^{+},$ $E_{res}=3.140(10)$ MeV, $\Gamma _{res}=68(6)$ keV, and $J^{\pi
}=2^{+},$ $E_{res}=5.123(11)$ MeV, $\Gamma _{res}=229(51)$ keV resonance
states, which are admissible and can be associated with $^{3}F_{2}$ wave. We
also consider the $M1$ transition for the $J^{\pi }=1^{+}$, $%
E_{res}=1.29(10) $ MeV of a non-resonance $^{3}P_{1}$ scattering wave to the
GS of $^{14}$O. Resonances with higher energies either have a large
momentum, or their momentum is not determined at all \cite{14} and are not
considered here.

\section{Interaction potentials}

\bigskip To find the radial wave functions $\chi _{i}$ and $\chi _{f}$ of the 
relative motion of particles in the initial and final states, respectively,
one should solve the Schr\"{o}dinger equation with potentials that describe
the $p^{13} $N scattering process and the states of the residual $^{14}$O
nucleus. The $p^{13}$N potentials for each partial wave, i.e., for the given
orbital angular momentum $L$ have a point-like Coulomb term, and a nuclear
part of the $p^{13}$N interaction. The nuclear part of potential can be
written in the one-range Gaussian form as \cite{3,NucPhys2015}

\begin{equation}
V(r,SLJ)=-V_{0}(SLJ)\exp (-\alpha _{SLJ}r^{2}),\text{ }  \label{Potential1}
\end{equation}%
where $r$ is the distance between the proton and $^{13}$N, $V_{0}(SLJ)$ is
the depth of the potential and $\alpha _{SLJ}$ is the range parameter for
given $S,$ $L,$ and $J,$ respectively. Resonance potentials were constructed
in such a way as to correctly describe the energy and width of such
resonances.

The interaction (\ref{Potential1}) is given as a two-parameter Gaussian
potential, i.e., with just an $LSJ$-dependent central term, and the 
consideration of Pauli-forbidden states is based on Young diagrams. Each
state is described independently, so the potential for each partial wave
effectively includes all features such as spin-orbit and spin-spin terms,
but without separation in operator terms. There are different approaches and
prescriptions related to the choice of the potential parametrization. In this
study we are using the one-range Gaussian potential (\ref{Potential1}),
which has only two fitting parameters, due to its simpler form than the
Woods-Saxon and also because at studies of the radiative capture processes
at low energies this potential allows complete description of all basic
characteristics of the process. Over 30 radiative capture reactions have
been successfully described (see \cite{3,NucPhys2015} and citations herein)
using the one-range Gaussian potential. One can also mention that a
comparison of studies of a radiative capture process using the Woods-Saxon
potential \cite{Alik47} and a simple one-range Gaussian potential \cite%
{Alik46} shows that the latter potential provided good description of the
process. Besides, the using of the Gaussian potential is easy due to the
fact that the expansion of WF in terms of the Gaussian basis within the
variation method \cite{3} the majority of matrix elements are obtained in
the close analytical form.
\begin{figure}[b]
\noindent
\begin{centering}
\includegraphics[width=10.0cm]{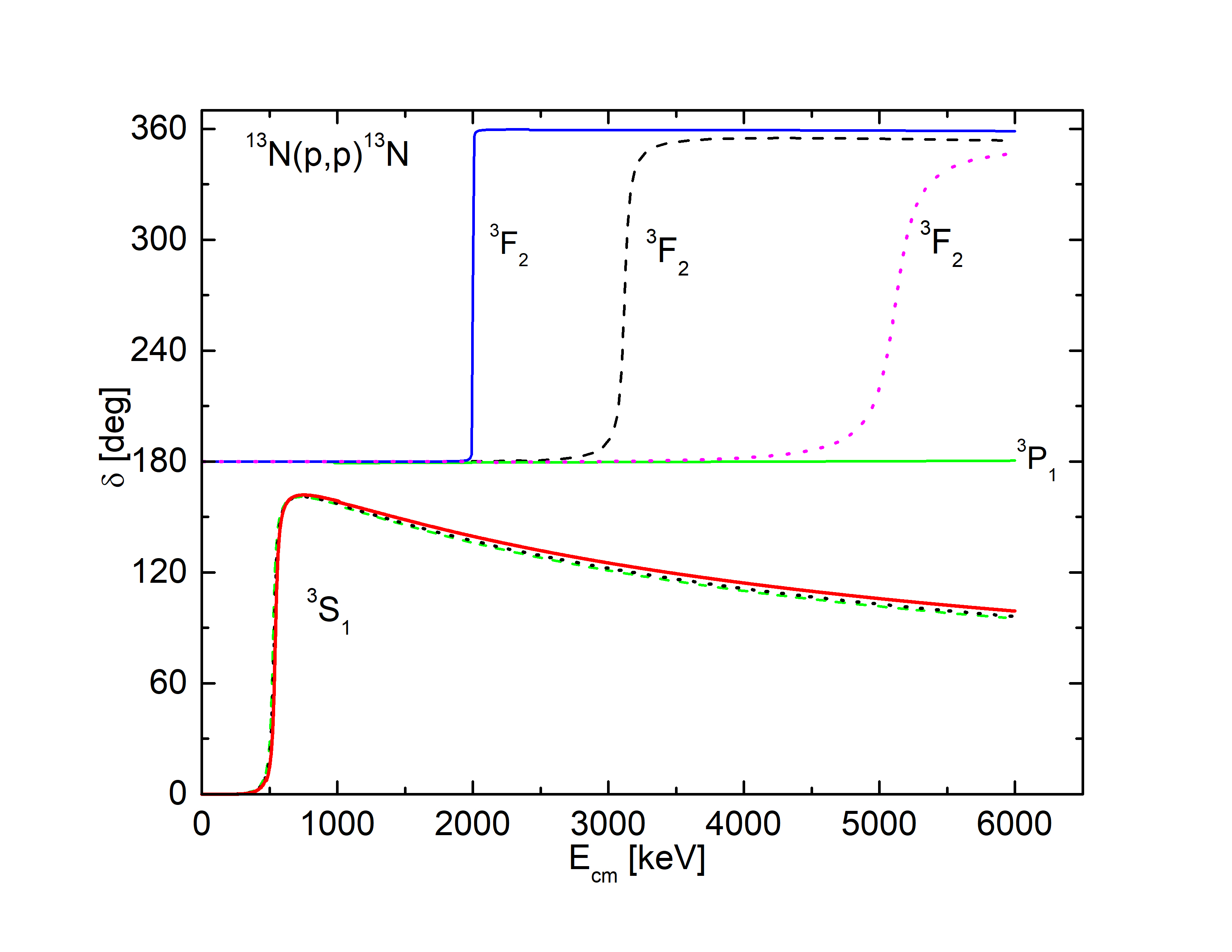}

\par\end{centering}
\caption{The dependence of the elastic $p^{13}N$ scattering phases on the
energy. Calculations are performed using the potentials with parameters from
Table \protect\ref{tab1}. The $^{3}S_{1}$ phase shift is calculated using
the set 1$a$ (green dashed curve), 1$b$ (black dotted curve) and 1$c$ (red
solid curve) from Table \protect\ref{tab1}, respectively. The three sets of
parameters for the potential give almost the coincide results for the $%
^{3}S_{1}$ phase shift. In the given energy region the $^{3}P_{1}$ phase
shift has very weak energy dependence. }
\label{fig1}
\end{figure}

In calculations we use for the proton mass $m_{p}=1.007276469$ amu \cite{8}
and $^{13}$N mass 13.0057367 amu \cite{9}, where 1 amu $=$ 931.4941024 MeV
\cite{8} and the constant $\hbar ^{2}/m_{0}=41.4686$ MeV$\cdot $fm$^{2}$.
The Coulomb potential at $R_{C}=0$ is written in MeV as $%
V_{C}(r)=1.439975Z_{1}Z_{2}/r$, where $r$ is the interparticle distance in
fm, $Z_{1}$ and $Z_{2}$ are charges of the particles in units of the
elementary charge. The Sommerfeld parameter $\eta =\mu
Z_{1}Z_{2}e^{2}/(k\hbar ^{2})=3.44476\text{\textperiodcentered }10^{-2}\mu
Z_{1}Z_{2}/k$, where $k=\left( 2\mu E/\hbar ^{2}\right) ^{1/2}$ is the wave
number specified in fm$^{-1}$ and defined by the energy $E$ of interacting
particles, and the reduced mass $\mu $ of these particles in amu.

Following Ref. \cite{Nichitiu1980} for calculations of the width employing
the resonance scattering phase we use the expression $\Gamma
_{res}=2(d\delta /dE)^{-1},$ where $\delta $ is the phase shift. For
description of $^{3}S_{1}$, $^{3}P_{1}$, and $^{3}F_{2}$ scattering states
we use the corresponding experimental energies and widths. For the $^{3}S_{1}
$ resonance there are reported three different experimental measurements for
the resonance energy and width. Therefore, we constructed the potential for
the $^{3}S_{1}$ resonance scattering phase with three sets of parameters. In
Table \ref{tab1} are given the results of calculations of parameters for the
corresponding potential. The potential with sets of parameters 1$a,$ 1$b,$
and 1$c$ reproduce the resonance energies 528, 536 and 545 keV,
respectively. The latter allows to find the optimal astrophysical $S$%
-factor. In Fig. \ref{fig1} the dependence of the elastic $p^{13}N$
scattering phase shifts on the energy $E_{cm}$. The result of calculation of
the $^{3}S_{1}$ phase shift with the set 1$c$ parameters for the $S$
scattering potential without FS leads to 90$^{0}\pm $1$^{0}$ at the energy $%
E_{res}=0.545$ MeV \cite{12} are presented by the red solid curve. The
calculations of the $^{3}S_{1}$ phase using the sets of parameters 1$a$ and 1%
$b$, which correspond to the resonances at $E_{res}=0.528$ MeV \cite{14} and
$E_{res}=0.536$ MeV \cite{15} give the coincide results in Fig. \ref{fig1}.
Thus, the scattering potentials with the set of parameters 1$a$, 1$b$ and 1$c
$ are phase shift equivalent potentials.

The potential of the nonresonance scattering is also constructed quite
unambiguously based on the scattering phase shifts for a given number of
bound states allowed and forbidden in the partial wave. The accuracy of
determining the parameters of such a potential is primarily associated with
the accuracy of extracting the scattering phase shifts from the experimental
data. Since the classification of states according to Young diagrams makes
it possible to unambiguously fix the bound states number, which completely
determines its depth, the potential width at a given depth is determined by
the shape of the scattering phase shift. When constructing a nonresonance
scattering potential from the data on the spectra of the nucleus, it is
difficult to evaluate the accuracy of finding its parameters even for a
given number of bound states. Such a potential, as is usually assumed for
the energy range up to $1-3$ MeV, should lead to the scattering phase shift
close to zero or gives a smoothly decreasing phase shift shape, since there
are no resonance levels in the spectra of the nucleus.

For the $^{3}P_{1}$ scattering potential, one can use the parameter set 2
from Table \ref{tab1}. Such a potential has the FS and leads to scattering
phase shift of 180$^{0}\pm $1$^{0}$, which has a very weak dependence of
energy and is presented by the green solid curve in the energy range from
zero to 7 MeV. Since it has the FS, according to the generalized Levinson
theorem, its phase shift begins at 180$^{0}$ \cite{7}.

\begin{table}[t]
\caption{List of transitions from the initial $\left\{ ^{\left( 2S+1\right)
}L_{J}\right\} _{i}$ state to $^{3}P_{0}$ GS of $^{14}$O nucleus. The value
of $P^{2}$ determines the coefficient in expressions (\protect\ref%
{Matelement}) and (\protect\ref{Matelement2}). The width $\Gamma _{res}$ and
$S(0)$-factor are obtained using the potential parameters $V_{0}$ and $%
\protect\alpha .$ The value $\protect\widetilde{S}(0)$ of the $S$-factor and
the set of parameters $1d$, $1e$, and $1f$ for the potential are used for
calculations of the resonance width $\protect\widetilde{\Gamma }_{res}.$ }
\label{tab1}
\begin{center}
\begin{tabular}{cccccccccccc}
\hline\hline
Set & \{$^{(2S+1)}L_{f}$\}$_{i}$ & Transition & $P^{2}$ & \multicolumn{2}{c}{%
$V_{0},$ MeV} & $\alpha ,$ fm$^{-2}$ & $E_{res},$ MeV & $\Gamma _{res},$ keV
& $S(0)$, keV$\cdot $b & $\widetilde{\Gamma }_{res}$, keV & $\widetilde{S}%
(0) $, keV$\cdot $b \\ \hline
&  &  &  & \textit{a} & 14.955 & 0.085 & 0.528(1) & 37(1) & 8.4(2) &  &  \\
&  &  &  & \textit{b} & 15.882 & 0.092 & 0.536(1) & 38(1) & 7.9(2) &  &  \\
1 & $^{3}S_{1}$ resonance at & $E1$ & 1 & \textit{c} & 18.244 & 0.11 &
0.545(1) & 37(1) & 7.0(2) &  &  \\
& $0.528,0.536,0.545$ MeV &  &  & \textit{d} & 35.053 & 0.25 & 0.528(1) &  &
& 22(1) & 4.8(1) \\
&  &  &  & \textit{e} & 29.316 & 0.02 & 0.536(1) &  &  & 25(1) & 5.1(1) \\
&  &  &  & \textit{f} & 31.582 & 0.22 & 0.545(1) &  &  & 26(1) & 4.9(1) \\
2 & $^{3}P_{1}$ no resonance & $M1$ & 2 & \multicolumn{2}{c}{555.0} & 1.0 &
&  & 0.014(1) &  &  \\
3 & $^{3}F_{2}$ resonance at 1.981(10) & $E2$ & 3 & \multicolumn{2}{c}{
698.134} & 0.36 & 2.000 & 13 & $<$ 0.01 &  &  \\
4 & $^{3}F_{2}$ resonance at 3.117(19) & $E2$ & 3 & \multicolumn{2}{c}{
343.613} & 0.18 & 3.120 & 58 & $<$ 0.01 &  &  \\
5 & $^{3}F_{2}$ resonance at 5.123(11) & $E2$ & 3 & \multicolumn{2}{c}{430.2}
& 0.23 & 5.127 & 232 & $<$ 0.01 &  &  \\ \hline\hline
\end{tabular}%
\end{center}
\end{table}

We also considered the $J^{\pi }=2^{+}$, $E_{res}=1.981(10)$ MeV, $\Gamma
_{res}=5$ keV, $J^{\pi }=2^{+},$ $E_{res}=3.140(10)$ MeV, $\Gamma
_{res}=68(6)$ keV, and $J^{\pi }=2^{+},$ $E_{res}=5.123(11)$ MeV, $\Gamma
_{res}=229(51)$ keV resonances, which lead to a noticeable change in the $S$%
-factor in resonance regions, using the potentials with the parameters set
3, 4 and 5, respectively, from Table \ref{tab1}. However, it was not
possible to construct such potentials in $P-$waves, therefore, $F$
scattering waves were used here. The first of them leads to a resonance at
2.00 MeV with a width $\Gamma _{res}=13$ keV shown by the blue solid curve
in Fig. \ref{fig1}, the second gives the resonance at $E_{res}=3.12$ MeV and
a width $\Gamma _{res}=58$ keV and is presented by the black dashed curve,
while the phase shift of the third resonance at $E_{res}=5.127$ MeV is shown
by the dotted curve. We were not able to obtain the resonance at $%
E_{res}=1.981$ MeV with the width $\Gamma _{res}<5$ keV, as given in \cite%
{14}, but the obtained value is completely consistent with the recent data
\cite{15}.

To build the potential for description of the GS of $^{14}$O, \ we use the
experimental binding energy and the asymptotic normalization coefficient ($%
ANC$)\ of this state. The corresponding potentials are tested based on the
calculation of the root mean square charge radius of $^{14}$O.


In Ref. \cite{16} the value of $ANC=5.42(48)$ fm$^{-1/2}$ and the proton
spectroscopic factor $S_{p}=1.88(34)$ are given. A similar value of $%
ANC=5.42(74)$ fm$^{-1/2}$ is also reported in Ref. \cite{17}, while Ref.
\cite{18} reports $ANC=5.39(38)$ fm$^{-1/2}$. Using the results of \cite{16}
for the $ANC$ and the expression for the asymptotic normalization constant
\begin{equation}
ANC=\sqrt{S_{p}}C
\end{equation}%
one gets $C=$4.04(72) fm$^{-1/2}$. For determination of $C$, the following
definition is also used (see, for example, \cite{19})
\begin{equation}
\chi _{L}(r)=CW_{-\eta ,L+1/2}(2k_{0}r),
\end{equation}%
where $W_{-\eta ,L+1/2}(2k_{0}r)$ is a Whittaker function. We use a
different definition of $ANC$ \cite{20}
\begin{equation}
\chi _{L}(r)=\sqrt{2k_{0}}C_{w}W_{-\eta ,L+1/2}(2k_{0}r)
\end{equation}%
which differs from the previous definition by the factor $\sqrt{2k_{0}}$
which in this case is 0.956. Then for the dimensionless $C_{w}$ we get $%
C_{w}=4.23(75)$. At the same time in Ref. \cite{18} $S_{p}=0.90(23)$ was
given for the spectroscopic factor, which yields $ANC=5.39(38)$ fm$^{-1/2}$
and allows to obtain $C_{w}=6.15(1.22)$. $ANC=30.4(7.1)$ fm$^{-1}$ and $%
S_{p}=1.94(45),$ were obtained in Ref. \cite{25}, which lead to the
dimensionless asymptotic normalization constant within the range 3.26 --
5.30 with an average of 4.28(1.02).

The potential of a bound ground $^{3}P_{0}$ state with the FS should
correctly reproduce the GS energy --4.628 MeV of $^{14}$O nucleus with $%
J^{\pi }=0^{+}$ in the $p^{13}$N channel \cite{12} and it is reasonable to
describe the mean square radius of $^{14}$O as well. Since data on the
radius of $^{14}$O are not available, we consider it to coincide with the
radius of $^{14}$N, the experimental value of which is 2.5582(70) fm \cite{9}%
. As a result, we obtained the following parameters for the GS potential,
which lead to $C_{w}=4.1(1)$:

\begin{equation}
V_{0}(1,1,0^{+})=226.230\text{ MeV, }\alpha (1,1,0^{+})=0.23\text{ fm}^{-2}.
\label{Pot1}
\end{equation}

The potential (\ref{Potential1}) with the parameters (\ref{Pot1}) gives for
the $^{14}$O nucleus the binding energy of 4.628 MeV and the root mean
square charge radius $R_{ch}=$ 2.55 fm. We used 0.8768(69) fm for the proton
radius \cite{8} and 2.4614(34) fm for the $^{13}$N radius. The latter radius
was taken to be the radius of $^{13}$C \cite{9}, because the $^{13}$N radius
is not available.

The GS potential which leads to $C_{w}=6.1(1)$ has parameters
\begin{equation}
V_{0}(1,1,0^{+})=156.728\text{ MeV, }\alpha (1,1,0^{+})=0.15\text{ fm}^{-2}.
\label{Pot2}
\end{equation}

The GS potential with parameters (\ref{Pot2}) gives a binding energy of
4.628 MeV and the root mean square charge radius $R_{ch}=$ 2.63 fm. One can
see that the potential (14) gives a larger radius than the potential (13),
so by simple estimates it is clear the GS with (14) should have larger cross
sections.

\begin{figure}[t]
\noindent
\begin{centering}
\includegraphics[width=8.5cm]{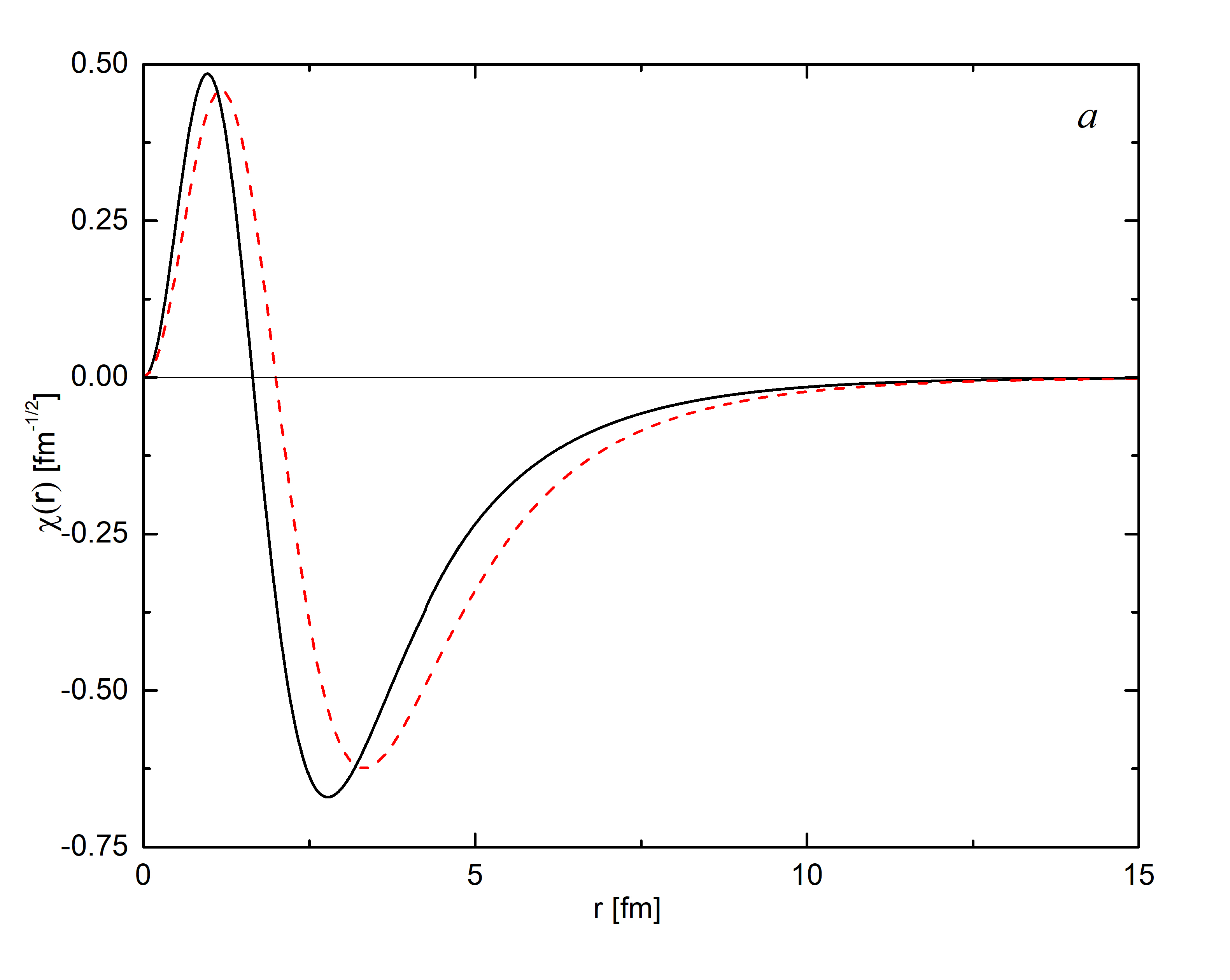}
\includegraphics[width=8.5cm]{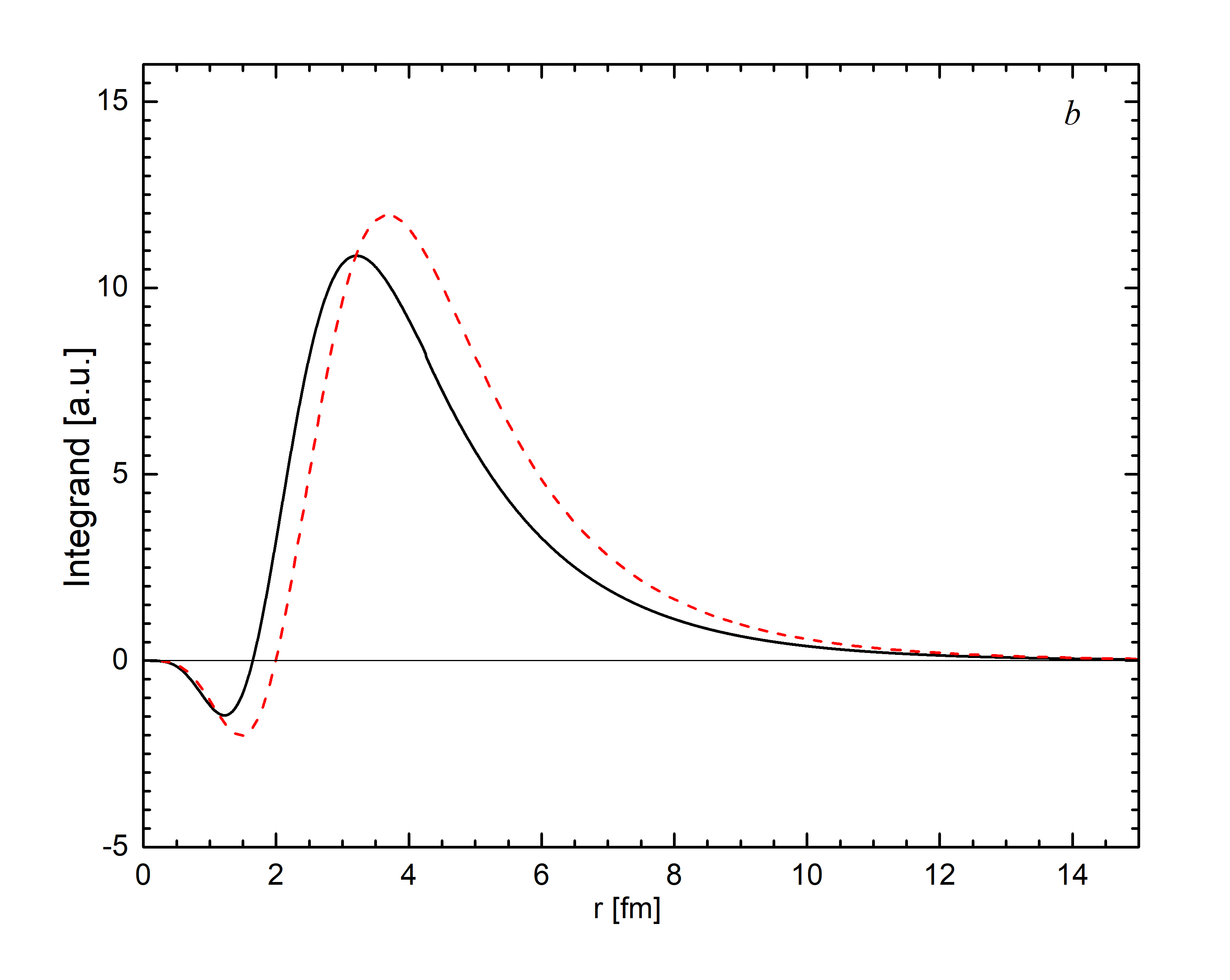}
\includegraphics[width=8.5cm]{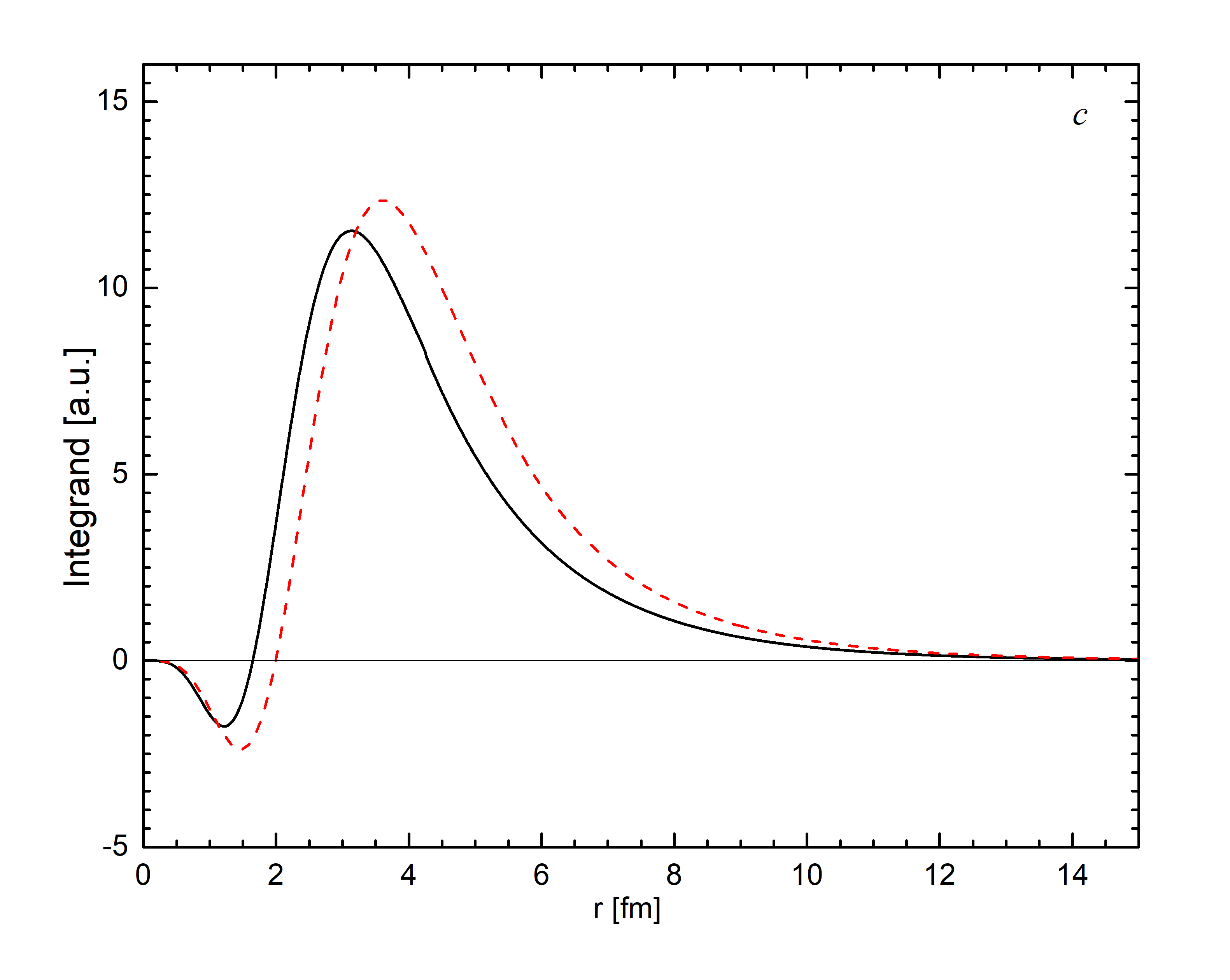}
\includegraphics[width=8.5cm]{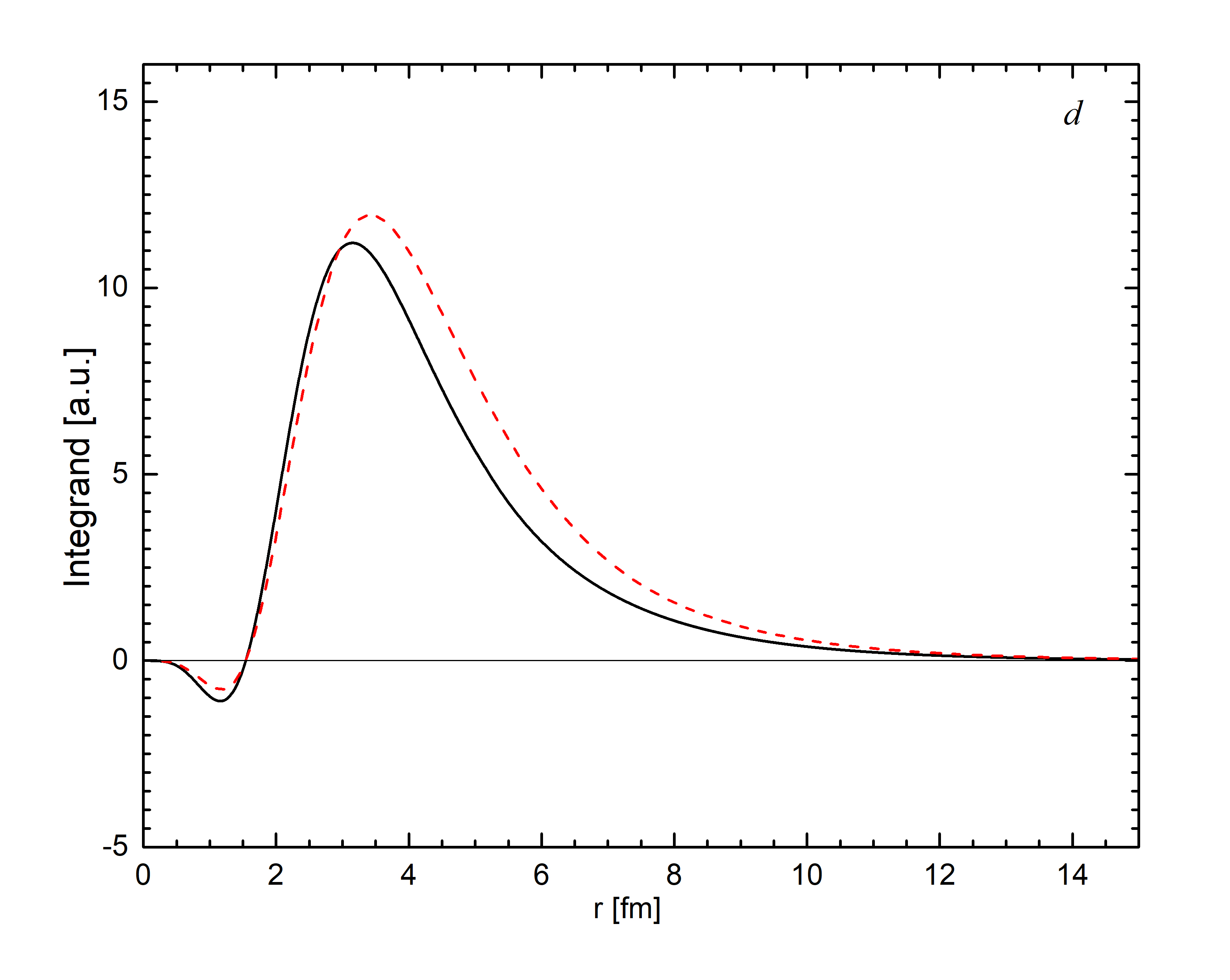}
\par
\end{centering}
\caption{(Color online) The radial part of the GS wave functions $^{14}$O in
the $p^{13}$N channel and integrands in the matrix element (\protect\ref%
{Overlapintegr}) for the $E1$ transition in arbitrary units (a.u.). (a) The
GS wave function obtained with potential (\protect\ref{Pot1}) (solid curve)
and potential (\protect\ref{Pot2}) (dashed curve). (b) The integrand of the $%
E1$ transition ME (\protect\ref{Overlapintegr}) for the scattering potential
with the set of parameters 1$a$ from Table I and for the GS potential (%
\protect\ref{Pot1}) (solid curve) and (\protect\ref{Pot2}) (dashed curve),
respectively. (c) The integrand of the $E1$ transition ME (\protect\ref%
{Overlapintegr}) for the scattering potential with the set of parameters 1$c$
from Table I and for the GS potential (\protect\ref{Pot1}) (solid curve) and
(\protect\ref{Pot2}) (dashed curve), respectively. (c) The integrand of the $%
E1$ transition ME (\protect\ref{Overlapintegr}) for the scattering potential
with the set of parameters 1$c$ from Table I and for the GS potential (%
\protect\ref{Pot1}) (solid curve) and (\protect\ref{Pot2}) (dashed curve),
respectively. (d) The integrand of the $E1$ transition ME (\protect\ref%
{Overlapintegr}) for the GS\ potential without FS (\protect\ref{Pot15}) at $%
C_{W}=4.1$ (solid curve) and the GS potential without FS (\protect\ref{Pot16}%
) at $C_{W}=6.0$ (dashed curve) and for the scattering potential (\protect
\ref{Pot17}), respectively.}
\label{fig3new}
\end{figure}

We calculated the radial WFs of GSs and shape of the integrand in matrix
element ME (\ref{Overlapintegr}) of the $E1$ transition using the scattering
potential with the set of parameters 1$a$ and 1$c$ from Table I. The results
of calculations are presented in Fig. \ref{fig3new}. The radial WFs for the
GS of $^{14}$O in the $p$$^{13}$N channel obtained with potentials (\ref%
{Pot2}) and (\ref{Pot2}) are shown in Fig. \ref{fig3new}$a$. The GS WFs have
the same behavior, different magnitudes and the shifted nodes. The different
magnitudes lead to the different shape of the integrand in the ME (\ref%
{Overlapintegr}) of the $E1$ transition, which also depends on the choice of
the parameters for the potential for the description of the scattering
state. The node in the nuclear interior leading to the node in the integrand
shown in Fig. \ref{fig3new}$b$ and \ref{fig3new}$c$, respectively. We should
be noted that integrands in the ME (\ref{Overlapintegr}) of the $E1$
transition almost coincide with the integrand shown in Fig. 8 in Ref. \cite%
{18}.

One should be noted that the shell model is undoubtedly the most perfectly
formulated from both a physical and mathematical point of view. In fact, on
the one hand, in the framework of shell model, the Pauli principle is
precisely taken into account. On the other hand, this model allows, based on
algebraic methods, to take into account the effects of clustering in atomic
nuclei. Thus, the shell model could be recognized as a criterion for testing
the ``quality'\ of other models using phenomenological nucleon-cluster
potentials. Let us for comparison consider the GS potentials without FS
and scattering potentials with the FS in the ${^3}S{_1}$ wave based on a
single-particle model. The GS potential without the FS has parameters:
\begin{equation}
V_{0}(1,1,0^{+})=61.23803\text{ MeV, }\alpha (1,1,0^{+})=0.13\text{ fm}^{-2}.
\label{Pot15}
\end{equation}
This potential leads to the binding energy of 4.62800 MeV, root-mean-square
charge radius $R_{ch} = 2.54$ fm and $C_{w} = 4.1(1)$. This completely
coincides with the option for potential (\ref{Pot1}). One can also obtain
another option for the GS potential, which agrees with the shell model of
the system, which has parameters:
\begin{equation}
V_{0}(1,1,0^{+})=45.46913\text{ MeV, }\alpha (1,1,0^{+})=0.085\text{ fm}%
^{-2}.  \label{Pot16}
\end{equation}
This potential leads to the binding energy of 4.62800 MeV, root-mean-square
charge radius $R_{ch} = 2.61$ fm and $C_{w} = 6.0(1)$. This coincides with
the option for potential (\ref{Pot2}). The scattering potential for the
resonance ${^3}S{_1}$ wave now has the FS and parameters:
\begin{equation}
V_{0}(1,1,0^{+})=125.529\text{ MeV, }\alpha (1,1,0^{+})=0.24\text{ fm}^{-2}.
\label{Pot17}
\end{equation}
This potential leads to the resonance energy of 545 keV and its width of
37(1) keV, this is completely coinciding with results for the set 1$c$ from
Table \ref{tab1}.
The shape of the integrands in the ME (\ref{Overlapintegr}) of the $E1$
transition for the GS potentials (\ref{Pot15}) and (\ref{Pot16}), and
scattering potential (\ref{Pot17}) is shown in Fig. \ref{fig3new}$d$.

We use the potentials with parameters from sets $1a$, $1b$ and $1c$ in Table %
\ref{tab1} for the description of the resonance states and parameters (\ref%
{Pot1}) and (\ref{Pot2}) for the description of the residual $^{14}$O
nucleus for calculations of the $^{13}$\textit{N(}$p,\gamma )^{14}$O
reaction rate and the astrophysical $S$-factor.

The astrophysical $S$-factor was calculated previously using the $^{3}S_{1}$
resonance scattering. 
Using the values of $\widetilde{S}(0)$ from Table \ref{tab1}, we consider
the inverse problem to construct potentials for description the $^{3}S_{1}$
resonance based on the resonance energies and the corresponding
astrophysical $S$-factor. The parameters of these potentials are given in
Table \ref{tab1} as sets $1d$, $1e$ and $1f$.

\section{Reaction rate and astrophysical $S$-factor of the proton radiative
capture on $^{13}$N}

Let us calculate the reaction rate for the $^{13}$N($p,\gamma $)$^{14}$O
radiative capture and the astrophysical $S$-factor using the total cross
section (\ref{Crossection}) and corresponding matrix elements of multipole
transition operators. The astrophysical factor $S(E)$ is defined as
\begin{equation}
S(E)=E\sigma _{c}(NJ,J_{f})e^{-2\pi \eta },  \label{Sfactor}
\end{equation}%
where the factor $\exp (-2\pi \eta )$ approximates the Coulomb barrier
between two point-like particles with charges $Z_{1}$ and $Z_{2}$ and
orbital momentum $L=0$, while for the reaction rate is commonly expressed in
cm$^{3}$mol$^{-1}$s$^{-1}$ and is determined according to Ref. \cite%
{28,Fowler} as
\begin{eqnarray}
N_{A}\left\langle \sigma _{c}v\right\rangle &=&N_{A}\frac{2(2/\pi )^{1/2}}{%
\mu ^{1/2}(k_{B}T)^{3/2}}\int_{0}^{\infty }\sigma _{c}(E)E\exp \left(
-E/k_{b}T\right) dE  \notag \\
&=&3.7313\times 10^{4}\mu ^{-1/2}T_{9}^{-3/2}\int_{0}^{\infty }\sigma
_{c}(E)E\exp \left( -11.605E/T_{9}\right) dE,  \label{Rate}
\end{eqnarray}%
where $N_{A}$ is Avogadro's number, $k_{B}$ is the Boltzmann's constant, $E$
is the energy in the center-of-mass frame given in MeV, the cross section $%
\sigma _{c}(E)$ is measured in $\mu $b, $\mu $ is the reduced mass in a.m.u,
and $T_{9}$ is the temperature in units of $10^{9}$ K. The behavior of $S$%
-factor, when resonances are present, in general, is expected to be rather
smooth at low energies and can be expanded in Taylor series around $E=0$
\cite{Baye2000,Ryan} as
\begin{equation}
S(E)=S_{0}+ES_{1}+E^{2}S_{2}.  \label{Sapproximation}
\end{equation}%
%
Essentially, the experimental data on the astrophysical $S$-factor of the
proton radiative capture on $^{13}$N are absent, but in the database \cite%
{21} there are rates of this reaction from Refs. \cite{16,22}. However, it
is clear that the shape of $S$-factor should mainly be determined by
resonance in the $^{3}S_{1}$ scattering wave at 0.528 MeV with a width $%
\Gamma _{res}=37.3(9)$ keV and $J^{\pi }=1^{-}$ \cite{14}. The contributions
of cross sections of $^{3}F_{2}$ resonances from Table \ref{tab1}, which are
determined by $E2$ transitions, are possible as well.

For calculations of the astrophysical $S$-factor we use the potentials with
parameters from sets $1a$, $1b$ and $1c$ in Table \ref{tab1} for the
description of the resonance state and parameters (\ref{Pot1}) and (\ref%
{Pot2}) for the description of the residual $^{14}$O nucleus. We also
calculate the width of $^{3}S_{1}$ resonance using the sets of the
parameters $1d$, $1e$ and $1f$ for the potentials from Table \ref{tab1},
which were obtained based on the values of the astrophysical $S$-factor.

\begin{figure}[h]
\noindent
\begin{centering}
\includegraphics[width=8.5cm]{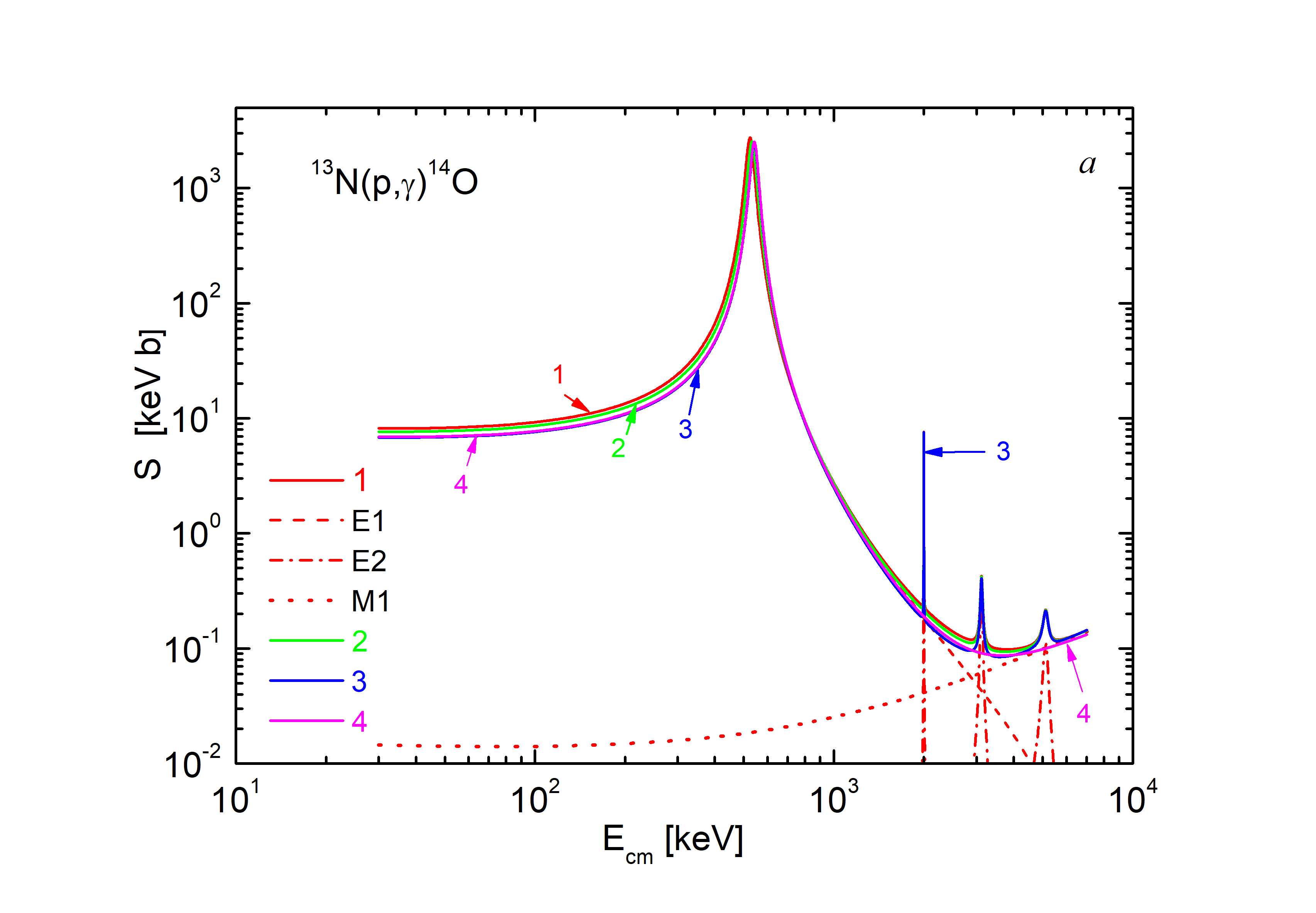}
\includegraphics[width=8.5cm]{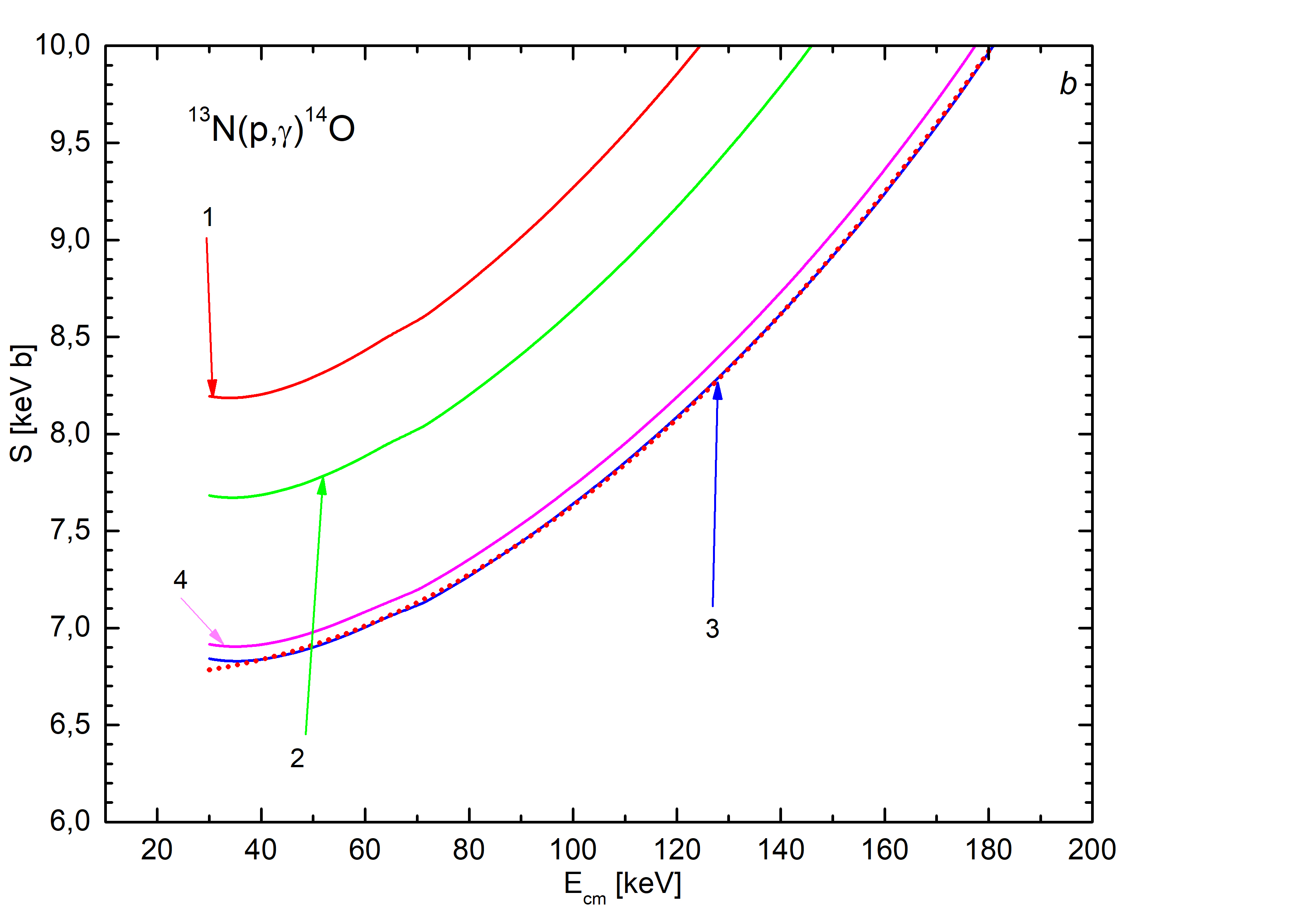}
\par
\end{centering}
\caption{(Color online) Astrophysical $S-$factor of the radiative proton
capture on $^{13}$N. ($a$) The energy range 30 keV -- 7 MeV. The solid
curves 1--4 present results of calculations which include the sum of $E1$, $%
E2$ and $M1$ transitions. Results presented by curves 1--3 are obtained
using potential with the set of parameters 1$a$, 1$b$ and 1$c$ from Table
\protect\ref{tab1}, respectively, and the GS potential (\protect\ref{Pot1}).
The curve 4 corresponds the "node inversion" in $E1$ transition simulated by
the GS potential without FS (\protect\ref{Pot15}) and scattering potential (%
\protect\ref{Pot17}) The dashed, dash-dotted and dotted curves illustrate
the contributions of the $E1$, $E2$ and $M1$ transitions, respectively, into
$S-$factor obtained for the potentials with the set of parameters 1$a$ from
Table \protect\ref{tab1} and GS (\protect\ref{Pot1}). ($b$) The energy range
30 -- 200 keV. The solid curves 1, 2, 3 and 4 present the same results as in
Fig. \protect\ref{fig2}$a$. The red dotted curve, which coincides with the
curve 3, presents the quadratic approximation (\protect\ref{Sapproximation})
of the $S$-factor at low energies.}
\label{fig2}
\end{figure}

The results of calculation of the $S$-factor of the radiative proton capture
on $^{13}$N to the GS of $^{14}$O nucleus include the sum of $E1$, $E2$ and $%
M1$ transitions are shown in Fig. \ref{fig2}. For the contribution of the $%
^{3}S_{1}$ scattering wave the set of parameters from Table \ref{tab1} for
the potential and potential (\ref{Pot1}) for the GS are considered. We
calculated the contributions of the $M1$ transition $^{3}P_{1}\rightarrow $ $%
^{3}P_{0},$ as well as the resonance $E2$ transitions into the $S$-factor
using the set of the potentials 2, 3, 4, and 5 from Table \ref{tab1},
respectively, and for the description of the GS\ the potential (\ref{Pot1})
was used. The results of these calculations are shown in Fig. \ref{fig2}$a$.
Analysis of results presented in Fig. \ref{fig2}$a$ shows that contributions
of the $M1$ and $E2$ transitions in the $S$-factor are negligible at
energies $E<1$ Mev, but are significant at high energies. At the resonance
energy, the $S$-factor reaches 2.4 MeV$\cdot $b, which is in good agreement
with the results of other works (see, for example, Refs. \cite{16,18,22, 23}%
), where the values for the $S$-factor from about 2.0 to 2.5
keV\textperiodcentered b were reported. The $S$-factor shown in Fig. \ref%
{fig2}$b$ is given for three sets of parameters $1a$, $1b$ and $1c,$
highlighting the differences. Results of our calculations for the $S$-factor
for the potentials $1a$ from Tables 1 and (\ref{Pot1}) in the energy range
of $30-50$ keV lie in the range of $8.2-8.3$ keV$\cdot $b, while in the
energy range of 30--70 keV, the average value is 8.4(2) keV$\cdot $b. The
error given here is determined by averaging $S$-factor over the above energy
range. Known results for the $S-$factor at zero energy lead to a value in
the range from 2.0 keV$\cdot $b to 6.0 keV$\cdot $b \cite{16,18,22,23}. We
use the GS potential (\ref{Pot2}) and calculate the $S$-factor in the energy
range $30-70$ keV using the set of parameters $1a$ from Table \ref{tab1} for
the potential and obtain almost constant value $S=11.9(2)$
keV\textperiodcentered b. At the resonance energy the $S$-factor reaches 2.9
MeV$\cdot $b, which is noticeably more than the results of \cite{16,18,22,23}%
. Therefore, we should mention that the GS potential with the parameters (%
\ref{Pot1}) for description of the GS of $^{14}$O nucleus in the $p^{13}$N
channel at low energy region leads to more preferable results for the
astrophysical $S$-factor, which are quite consistent with results from
previous calculations. Our calculations for the $S$-factor with the
parameters (\ref{Pot2}) for the potential of the GS gives a too high value
for the $S$-factor at low energies. However, since there are no experimental
measurements of the $S$-factor for this reaction, no final conclusions can
be drawn.

\begin{table}[b]
\caption{Astrophysical $S$-factors at zero energy.}
\label{tab2}
\begin{center}
\begin{tabular}{cccccc}
\hline\hline
Refs. &  & \cite{24} & \cite{16,17,18,23,25}$^{a}$ & \ \cite{26}$^{b}$ & [27]%
$^{a}$ \\ \hline
$S$, keV$\cdot b$ &  & $3.8_{-08}^{+1}$ & $5-6$ & $2.6$ & $2-2.3$ \\ \hline
\multicolumn{6}{c}{} \\
\multicolumn{6}{c}{$^{a}$Values are taken from Figures in Refs: [16] -- Fig.
7; [17] -- Fig. 8;} \\
\multicolumn{6}{c}{\lbrack 18] -- Fig. 9; [23] -- Fig. 5; [25] -- Fig. 3;
[27] -- Fig. 2b.} \\
\multicolumn{6}{c}{$^{b}$Value is taken from the approximation at low
energies.} \\ \hline\hline
\end{tabular}%
\end{center}
\end{table}

Table \ref{tab2} displays the compilation of the results for the astrophysical $%
S$-factors at zero energy obtained in different works. As can be seen
from Table \ref{tab2}, the deviation of data for the $S$-factor is in the
range from 2 to 6 keV\textperiodcentered b, although the most recent value
is apparently given in Ref. \cite{24}. We use the sets of parameters $1a$, $%
1b$, and $1c$ for the potential of $^{3}S_{1}$ scattering from Table \ref%
{tab1} and potential (\ref{Pot1}) for the GS, which reproduce accurately the
position and width of resonances and calculated corresponding $S$-factors.
The results are presented in Table \ref{tab1}. Depending on the resonance
energy $S$-factors are: 8.4(2) keV$\cdot $b ($E_{res}=528(1)$ keV), 7.9(2)
keV$\cdot $b ($E_{res}=536(1)$ keV), and 7.0 keV$\cdot $b ($E_{res}=545(1)$
keV). The potential with the set $1a$ from Table \ref{tab1} accurately
reproduces the width average value of 37 keV \cite{14} and leads to $%
S(0)=8.4(2)$ keV$\cdot $b. The potential with the set $1b$ reproduces the
resonance energy of 536(12) keV and the width $\Gamma _{res}=38(2)$ keV from
Ref. \cite{15}. The corresponding average value for the $S$-factor at 30--70
keV is $S(0)=7.9(2)$ keV$\cdot $b, which is slightly less than for the $S$
scattering potential $1a.$ We consider a potential with parameters $1c$,
which leads to the resonance at 545 keV and a width $\Gamma _{res}=37(1) $
keV \cite{12}. This potential gives $S(0)=7.0(2)$ keV$\cdot $b.

\begin{figure}[t]
\noindent
\begin{centering}
\includegraphics[width=9cm]{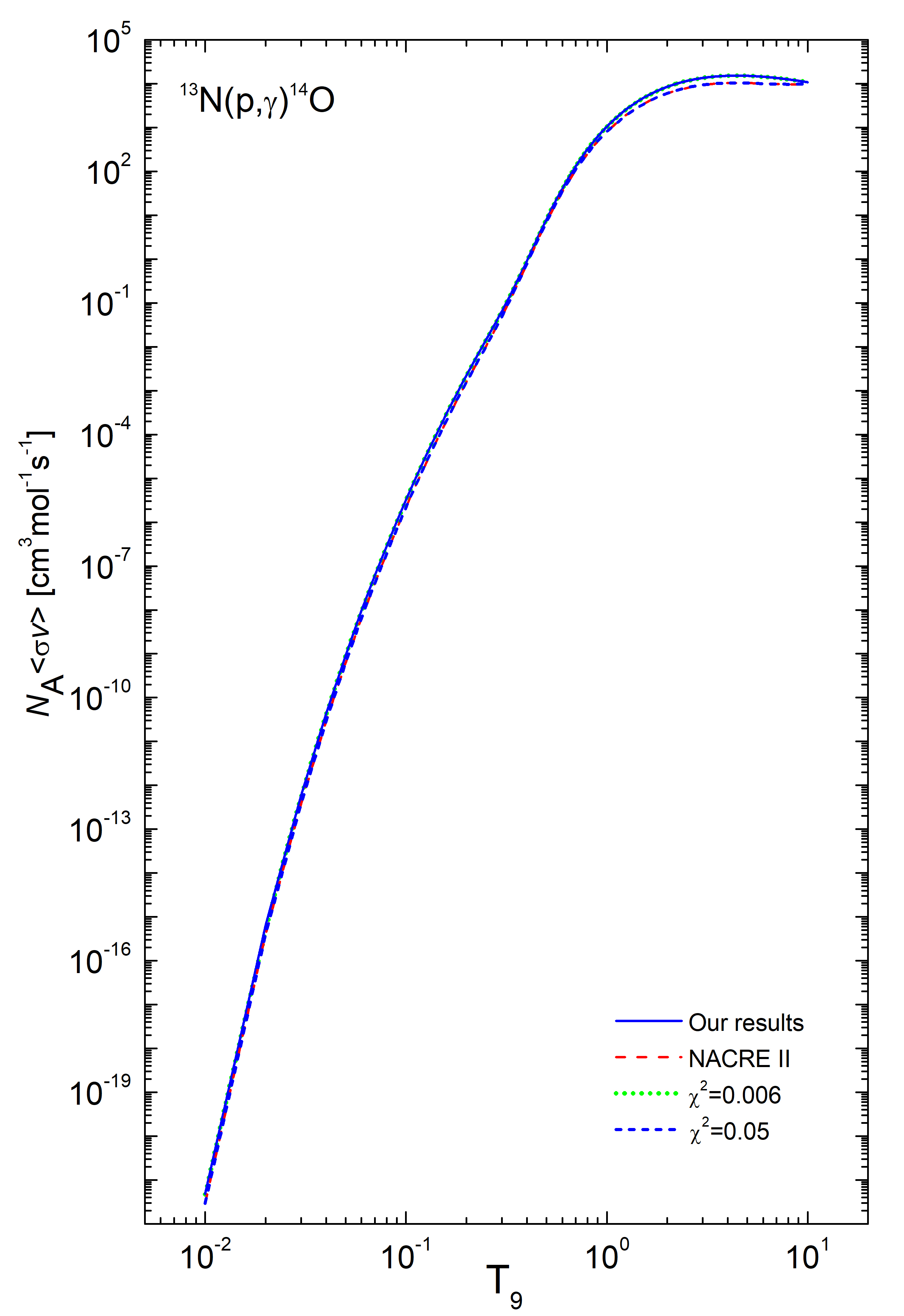}
\par\end{centering}
\caption{(Color online) The dependence of the reaction rate of $p^{13}N$
radiative capture on astrophysical temperature. The solid curve presents our
calculations for the sum of $E$1, $M1$ and $E2$ transitions performed for
the potentials with the set of parameters 1$c$ from Table (\protect\ref{tab1}%
) and GS (\protect\ref{Pot1}). Results of parametrization by Eq. (21)(%
\protect\ref{Param5}) with parameters from Table \protect\ref{tab4} for the
present calculations and the NACRE II data \protect\cite{24} are shown by
the dotted and dashed curves, respectively. The short dashed curve shows the
calculations using approximation (\protect\ref{Sapproximation}).}
\label{fig3}
\end{figure}

\begin{table}[b]
\caption{The results of the dependence of the $p^{13}N$ reaction rate on
temperature.}
\label{tab3}
\begin{center}
\begin{tabular}{cc|cc}
\hline\hline
Temperature, $T_{9}$ & Reaction rate, cm$^{3}$mol$^{-1}$s$^{-1}$ &
Temperature, $T_{9}$ & Reaction rate, cm$^{3}$mol$^{-1}$s$^{-1}$ \\ \hline
0.01 & 4.81E-22 & 0.35 & 2.53E-01 \\
0.02 & 6.46E-16 & 0.4 & 9.10E-01 \\
0.03 & 5.94E-13 & 0.45 & 2.91E+00 \\
0.04 & 4.37E-11 & 0.5 & 8.04E+00 \\
0.05 & 9.28E-10 & 0.6 & 4.02E+01 \\
0.06 & 9.54E-09 & 0.7 & 1.30E+02 \\
0.07 & 6.14E-08 & 0.8 & 3.13E+02 \\
0.08 & 2.86E-07 & 0.9 & 6.13E+02 \\
0.09 & 1.05E-06 & 1 & 1.04E+03 \\
0.1 & 3.22E-06 & 1.5 & 4.53E+02 \\
0.11 & 8.61E-06 & 2 & 8.46E+03 \\
0.12 & 2.06E-05 & 2.5 & 1.15E+04 \\
0.13 & 4.49E-5 & 3 & 1.35E+04 \\
0.14 & 9.09E-05 & 3.5 & 1.46E+04 \\
0.15 & 1.73E-04 & 4 & 1.51E+04 \\
0.16 & 3.12E-04 & 4.5 & 1.52E+04 \\
0.17 & 5.37E-04 & 5 & 1.51E+04 \\
0.18 & 8.90E-04 & 6 & 1.44E+04 \\
0.19 & 1.42E-03 & 7 & 1.35E+04 \\
0.2 & 2.21E-03 & 8 & 1.25E+04 \\
0.25 & 1.40E-02 & 9 & 1.16E+04 \\
0.3 & 6.46E-02 & 10 & 1.08E+04 \\ \hline\hline
\end{tabular}%
\end{center}
\end{table}

\begin{table}[t]
\caption{Parameters of the analytical parametrization of the $^{13}$N($p,%
\protect\gamma $)$^{14}$O reaction rate for the present calculations based
on Eq. (\protect\ref{Param5}) and NACRE II data \protect\cite{24} based on
Eq. (\protect\ref{Param5}) as well.}
\label{tab4}
\begin{center}
\begin{tabular}{c|ccccccc}
\hline\hline
Parameters & $a_{1}$ & $a_{2}$ & $a_{3}$ & $a_{4}$ & $a_{5}$ & $a_{6}$ & $%
a_{7}$ \\ \hline
Present work, Eq. (\ref{Param5}) & $4.68425$ & $5.5271$ & $72207.8$ & $%
-2.86832$ & $-17716.6$ & $-1304.726$ & $-1155.274$ \\
NACRE II & $77.14845$ & $4.87776$ & $-2791.957$ & $7554.465$ & $-4686.978$ &
$3691.79$ & $-4033.686$ \\ \hline\hline
Parameters & $a_{8}$ & $a_{9}$ & $a_{10}$ & $a_{11}$ & $a_{12}$ & $a_{13}$ &
$a_{14}$ \\ \hline
Present work, Eq. (\ref{Param5}) & $-1020.536$ & $215.4007$ & $4.66187\times
10^{6}$ & $10.92388$ & $8.5529\times 10^{7}$ & $15.50687$ & $16674.76$ \\
NACRE II & $1901.048$ & $-309.4704$ & $-3320.309$ & $7.12181$ & $%
3.13709\times 10^{8}$ & $15.87507$ & $-13.31191$ \\ \hline\hline
Parameters &  &  & $a_{15}$ & $a_{16}$ & $a_{17}$ &  &  \\ \hline
Present work, Eq. (\ref{Param5}) &  &  & $7.86955$ & $-77.74082$ & $1.38331$
&  &  \\
NACRE II &  &  & $5.65906$ & $-48.07274$ & $1.23332$ &  &  \\ \hline\hline
\end{tabular}%
\end{center}
\end{table}
Nevertheless, let us try to find out whether it is possible within our
approach to obtain the $S$-factor at zero energy that is close to the
results of \cite{24}, namely, $3.8_{-08}^{+1}$ keV\textperiodcentered b. We
constructed $S-$wave scattering potentials, which with the potential (\ref%
{Pot1}) for the GS, allow us to obtain maximum value of the $S-$factor about
$4.8-5.0$ keV$\cdot $b given in Ref. \cite{24}. Such potentials have the set
of parameters $1d,$ $1e,$ and $1f$ listed in Table \ref{tab1}. These
potentials lead to the resonance energies $528(1)$ keV, $536(1)$ keV, and
545(1) keV, respectively, but the corresponding widths are significantly
smaller than reported in Refs. \cite{14,15,12}. In particular, the set $1d$
leads to $E_{res}=528(1)$ keV, but the width is $\widetilde{\Gamma }%
_{res}=22(1)$ keV. At 30 keV $\widetilde{S}(0)=$4.8 keV$\cdot $b and its
average value in the range of $30-70$ keV is $\widetilde{S}(0)=4.8(1)$ keV$%
\cdot $b. If for the potential with a resonance energy of 536 keV, we use
the parameters $1e$ from Table \ref{tab1}, which lead to $\widetilde{\Gamma }%
_{res}=25(1)$ keV, then the $S$-factor decreases to $\widetilde{S}(0)=5.1(1)$
keV$\cdot $b. The $S$-factor decreases to $\widetilde{S}(0)=4.9(1)$ keV$%
\cdot $b, when we use the set $1f$ for the potential and the width becomes $%
\widetilde{\Gamma }_{res}=26(1)$ keV. Thus, in principle, all previously
obtained results for the $S$-factor at zero energy can be reproduced, but
the width of the resonances does not correspond to the data \cite{14,15,12}.
Therefore, for the considered resonance energies, if we correctly describe
their widths, it is impossible to obtain the $S$-factor below 7.0 (2) keV$\cdot $%
b. Only a decrease in the resonance width to 25--26 keV with its energy of
536--545 keV leads to the $S$-factor of the order of 4.9--5.1 keV$\cdot $b.

We also calculated the $S$-factor using the GS potential (\ref{Pot15})
without FS and the scattering potential (\ref{Pot17}). The result for the
average value of the $S$-factor in the range of 30-70 keV is 7.0(1) keV$%
\cdot $b that completely coincides with the $S$-factor, calculated with the
parameters set 1$c$ from Table \ref{tab1} and GS potential (\ref{Pot1}). We
use Eq. (\ref{Sapproximation}) for the approximation of the $S-$factor at
low energies. The corresponding parameters are: $S_{0}=6.7645$, $%
S_{1}=-2.7612\times 10^{-3}$, $S_{2}=1.1428\times 10^{-4}$ at $\chi
^{2}=1.0\times 10^{-3}$. The results are shown in Fig. \ref{fig2}$b$ by the
dotted curve that coincides with the curve 3, which presents the results of
calculations for the potentials with the set of parameters 1$c$ from Table %
\ref{tab1} and GS (\ref{Pot1}).

Using Eq. (\ref{Rate}), we calculated the rate of the $^{13}$N($p,\gamma $)$%
^{14}$O radiative capture by considering the sum of $E$1, $M1$ and $E2$
transitions. The dependence of the $^{13}$N($p,\gamma $)$^{14}$O reaction
rate on astrophysical temperature is shown in Fig. \ref{fig3}. The
corresponding rates are tabulated in Table \ref{tab3} for $0.01<T_{9}<10$.
The calculations are performed using the set of parameters $1c$ and (\ref%
{Pot1}) for the potentials. Let us mention that the earlier calculations
\cite{16,18,23} practically coincide with our results with small deviations,
while results from Ref. \cite{22} at temperatures $T_{9}>1$ are up to 2
times lower than present results. The results of calculations with the set
of parameters $1c$ and (\ref{Pot2}) for the potentials give a noticeable
excess of the reaction rate over the rates obtained with the GS potential (%
\ref{Pot1}) at temperatures above 1 $T_{9}.$

Following Ref. \cite{29} the reaction rate obtained in our calculations is
parameterized as

\begin{eqnarray}
N_{A}\left\langle \sigma v\right\rangle &=&\frac{a_{1}}{T}\exp \left( -\frac{%
a_{2}}{T}\right) \left(
1+a_{3}T^{1/3}+a_{4}T^{2/3}+a_{5}T^{4/3}+a_{7}T^{5/3}+a_{8}T^{6/3}+a_{9}T^{7/3}\right) +
\notag \\
&&\frac{a_{10}}{T^{1/2}}\exp \left( -\frac{a_{11}}{T^{1/2}}\right) +\frac{%
a_{12}}{T}\exp \left( -\frac{a_{13}}{T^{1/3}}\right) +\frac{a_{14}}{T^{1/3}}%
\exp \left( -\frac{a_{15}}{T^{1/2}}\right) +\frac{a_{16}}{T^{2}}\exp \left( -%
\frac{a_{17}}{T^{2}}\right) .  \label{Param5}
\end{eqnarray}

The parameters for the reaction rate (\ref{Param5}) from Table \ref{tab4}
lead to $\chi ^{2}=0.006$, and allow to merge with the calculated reaction
rate using Eq. (\ref{Param5}). Results of calculations using Eq. (\ref%
{Param5}) are presented in Fig. \ref{fig3}. It almost merges with a blue
solid curve that shows the calculated reaction rate using Eq. (\ref{Rate})
that is given in Table \ref{tab3}. We parameterized the NACRE\ II data \cite%
{24} using the same Eq. (\ref{Param5}) with $\chi ^{2}=0.05$ and 5\% errors,
which leads to the parameters listed in Table \ref{tab4}. The corresponding
results of calculations are shown in Fig. \ref{fig3} by the dashed curve.

\begin{figure}[t]
\begin{centering}
\includegraphics[width=8.5cm]{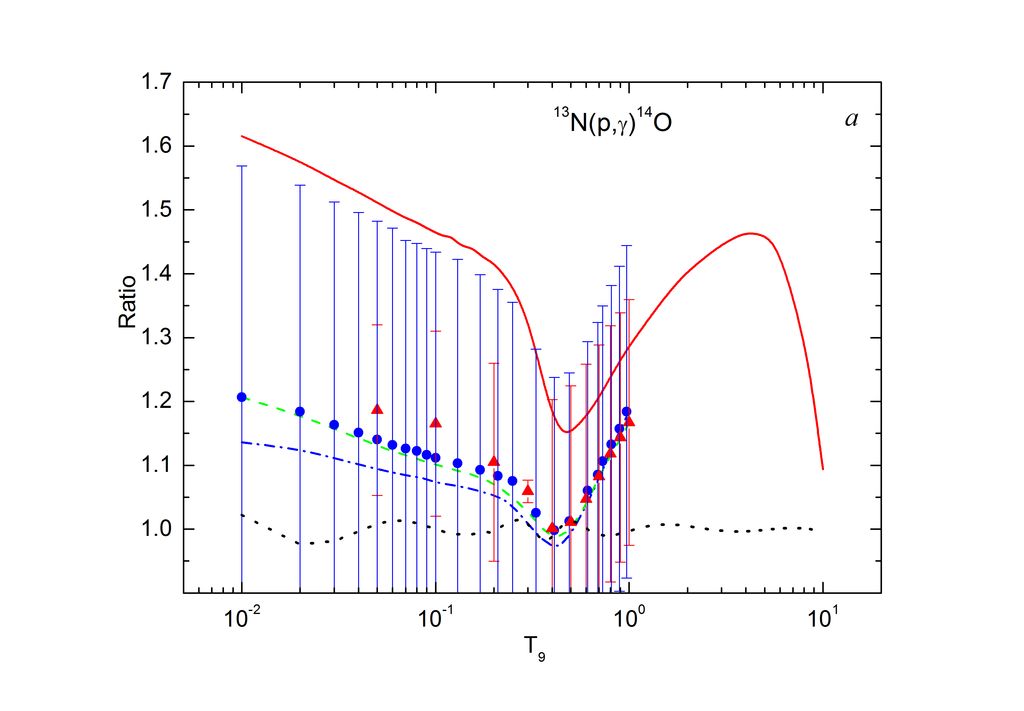}
\includegraphics[width=8.5cm]{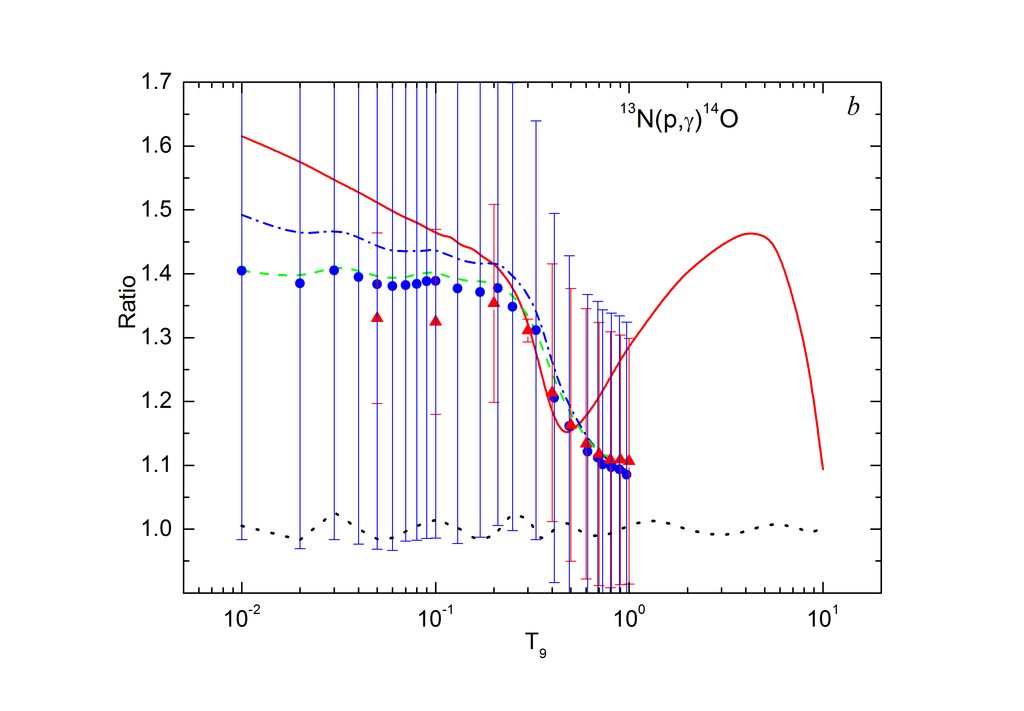}
\par\end{centering}
\caption{(Color online) The dependence of the ratio of the reaction rates on
temperature. \$a\$) The ratio of the reaction rate obtained in the present
calculations and given in Refs. \protect\cite{16,18,23,24,25},
correspondingly : NACRE II \protect\cite{24} - solid curve, Li, et al.
\protect\cite{16} - dash-dotted curve, Guo, et al. \protect\cite{27} -
dashed curve, Tang, et al. \protect\cite{18} - circles with error bars,
Magnus, et al. \protect\cite{23} - triangles with error bars, the dotted
curve is the ratio of the estimated speed to its approximation (\protect\ref%
{Param5}); \$b\$) The ratio of the reaction rates obtained in the present
calculations, Refs. \protect\cite{16,18,23,25} to the NACRE II \protect\cite%
{24} approximated with parameters from Table \protect\ref{tab4}: present
calculations - solid curve, Li, et al. \protect\cite{16} - dashed curve,
Guo, et al. \protect\cite{27} - dash-dotted curve, Tang, et al. \protect\cite%
{18} - circles with error bars, Magnus, et al. \protect\cite{23} - triangles
with error bars, correspondingly. The dotted curve corresponds to the ratio
of NACRE II data and its approximation using the parameters from Table
\protect\ref{tab4}. }
\label{fig4}
\end{figure}

For the detailed comparison of the dependence of the reaction rate on
astrophysical temperature, we calculated the ratio of our reaction rate to
the rates from Refs. \cite{24, 16,25,18,23}. The results of this comparison
are shown in Fig. \ref{fig4}$a.$ It can be seen from Fig. \ref{fig4}$a$ that
the results of present calculations exceed NACRE II up to 1.7 times at the
lowest temperatures and are almost equal to them at a temperature of $10$ $%
T_{9}$. The results of other studies lead to values that go below present
calculations up to 1.2 times at a temperature of 0.01 $T_{9}$, and in the
range of $0.4-0.5$ $T_{9}$ practically coincide with our data. But as the
temperature tends to 1 $T_{9}$, the values again become less than ours by
1.2 times. In Fig. \ref{fig4}$b$ are presented the ratios of the reaction
rates obtained in the present work and in Refs. \cite{16,25,18,23} to the
NACRE II \cite{24} which is parameterized with the parameters from Table \ref%
{tab4}.

Let us make a comparative analysis for the $S$-factor obtained within our
approach and calculated in the $R$-matrix approach \cite{18,181}. Ref. \cite%
{18} presents the most detailed and accurate uncertainties analysis for the
astrophysical $S$-factor, where the uncertainties were investigated by
varying 5 parameters: the $ANC$ for $^{14}$O, $\Gamma _{\gamma }$, $\Gamma
_{tot}$, and $E_{c.m}$ of the first resonance. The authors concluded that
with increasing energy, the fractional uncertainty in the $S$-factor drops
from 0.31 to 0.21 and the uncertainty of the $\Gamma _{\gamma }$ and the
total width of the first resonance $\Gamma _{tot}$ as well as the $ANC$ make
significant contributions to the uncertainty for $E_{cm}<0.6$ MeV \cite{18}.

In our model we operate with 3 experimental input parameters, i.e. $%
ANC $, $\Gamma _{tot}$, and $E_{cm}$. So, the initial score is $5:3$. The
uncertainty of $E_{cm}$ only produces less than a 2\% \cite{18}. Therefore,
it is reasonable to exclude the $E_{cm}$ among both parameter sets as the
consensus holds. Thus, the score drops to $4:2$. The $\Gamma _{\gamma }$
rises the highest uncertainty -- 20$-$30\% \cite{18}.

In our model there is no such uncertainty because we do not subdivide the
capture cross section into a direct and resonant parts and we operate 
with $ANC$ and $\Gamma _{tot}$ only. The signature of the resonances is seen
in phase shifts energy dependence shown in Fig. \ref{fig1}. In our
calculations the resonances are incorporated in natural continuous form
without any subdivisions. So that, there is no need for the $\Gamma _{\gamma
}$ parameter. Also, it is important to mention that we are implementing the
calculations of the overlap integrals starting from $r=0$, contrary to \cite%
{16,17,18,25}, where the channel radius cut-off parameter is exploited.
Concerning to the $ANC$: we examined the cases with $ANC_{min}$ and $%
ANC_{max}$ and found $ANC_{opt}$, within the correlation of $\Gamma _{tot}$.
Results in \cite{16,17,18,25} are obtained based on the averaged $ANC$, and
did not examine or show the band variety on the cross sections or $S$%
-factors within this very context.

\section{CNO AND HOT CNO CYCLES}

Since the late 1930s, when von Weizs\"{a}cker \cite{Weizsacker} and Bethe
\cite{Bethe} independently proposed sets of fusion reactions by which stars
convert hydrogen to helium, it is well established that the
carbon-nitrogen-oxygen cycles is a mechanism for hydrogen burning in stars.
The dominant sequence of reactions for this cycle is the following

\begin{equation}
^{12}\text{C}(p,\gamma )^{13}\text{N(}e^{+}\nu _{e})^{13}\text{C}(p,\gamma
)^{14}\text{N(}p,\gamma )^{15}\text{O(}e^{+}\nu _{e})^{15}\text{N(}p,\alpha
)^{12}\text{C.}  \label{CNC}
\end{equation}%
The character of the nuclear burning is extremely temperature sensitive and,
when temperature is low enough, the hot carbon-nitrogen-oxygen
cycle\textit{\ }

\begin{equation}
^{12}\text{C}(p,\gamma )^{13}\text{N(}p,\gamma )^{14}\text{O}(e^{+}\nu
_{e})^{14}\text{N(}p,\gamma )^{15}\text{O(}e^{+}\nu _{e})^{15}\text{N(}%
p,\alpha )^{12}\text{C}  \label{HCNC}
\end{equation}%
starts. Since, at low $T_{9}$ temperatures the $^{13}$N($p,\gamma )^{14}$O
reaction in the sequence (\ref{HCNC}) is competitive with the $^{13}$N($%
e^{+}\nu _{e})^{13}$C decay in the sequence (\ref{CNC}), the formation and
decay of $^{14}$O becomes a major distinguishing feature of this higher
temperature cycle. Therefore, the stellar $^{13}$N($p,\gamma )^{14}$O
reaction rate determines the order and the precise temperature of the
conversion of the cold CNO cycle to the HCNO cycle and the waiting point in
the cycle changes from $^{14}$N to the $^{14}$O and $^{15}$O and the $^{13}$%
N($p,\gamma )^{14}$O reaction is a key process which determines this
conversion.

One can say that the topic is hardly new, which is illustrated by the number
of references on the $S$-factor of the $^{13}$N($p,\gamma $)$^{14}$O
reaction and the different reactions rates \cite{16,17,23,26,18,25}. In Ref.
\cite{Smith} it was suggested the most consistent and accurate methodology for
analyses of the temperature and density conditions for the HCNO cycle. Below
we use this methodology along with our results for the $^{13}$N($p,\gamma $)$%
^{14}$O reaction rate and reanalyze the dependence of the lifetime against
hydrogen burning via $^{13}$N($p,\gamma $)$^{14}$O reaction as a function of
temperature and find the temperature window and densities of a stellar
medium at which the CNO cycle is converted to the hot CNO cycle. The
reanalysis is extended for the stellar density dependence on temperature.
Therefore, we use our results for the $^{13}$N($p,\gamma )^{14}$O reaction
rate, follow Ref. \cite{Smith} and find the temperature window and densities
of a stellar medium at which the CNO cycle is converted to the hot CNO
cycle. We can achieve the latter by comparing the $^{13}$N($p,\gamma )^{14}$%
O, $^{14}$N($p,\gamma )^{15}$O and $^{12}$C$(p,\gamma )^{13}$N reaction
rates and the lifetime of nuclei against destruction by hydrogen burning.

\begin{figure}[t]
\noindent
\begin{centering}
\includegraphics[width=11.5cm]{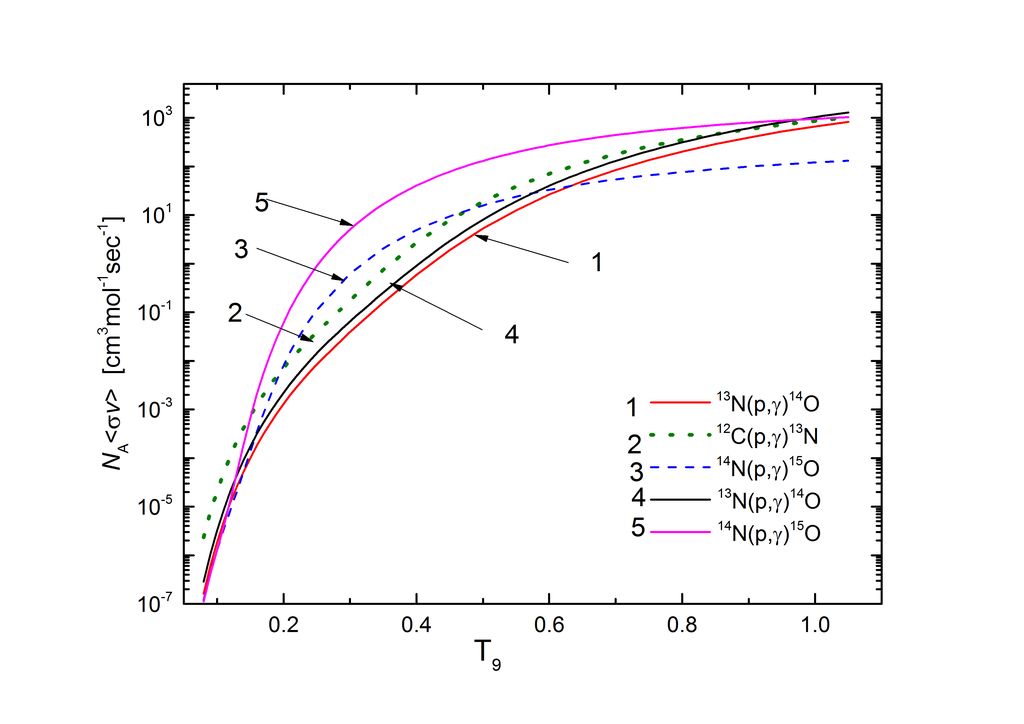}

\par\end{centering}
\caption{(Color online) The dependence of the reaction rates $%
N_{A}\left\langle \protect\sigma _{c}v\right\rangle $ on temperature for the
$^{12}$C$(p,\protect\gamma )^{13}$N, $^{13}$N($p,\protect\gamma )^{14}$O,
and $^{14}$N($p,\protect\gamma )^{15}$O reactions. Curves: 1, 2 and 3 - the
data are taken from Ref. \protect\cite{Smith}, 4 - present calculation, 5 -
results from Ref. \protect\cite{Dubovichenko2020}.}
\label{fig5}
\end{figure}

The lifetime of isotopes in the stellar CNO cycle relative to the combustion
of hydrogen one can determine as follows \cite{Rolfs1988,Ryan}

\begin{equation}
\tau =\frac{A_{H}}{\rho X_{H}}\frac{1}{N_{A}\left\langle \sigma
_{c}v\right\rangle },  \label{Time}
\end{equation}%
where $A_{H}$ is the atomic mass of hydrogen, $X_{H}$ is the relative
abundance of hydrogen by mass, $\rho $ is the density of the stellar medium,
and $N_{A}\left\langle \sigma _{c}v\right\rangle $ is the appropriate
proton-capture reaction rate. Thus, as it is follows from Eq. (\ref{Time}),
lifetime is determined precisely by the rate of the corresponding reaction.
In our calculations we use the $^{12}$C$(p,\gamma )^{13}$N, $^{13}$N($%
p,\gamma )^{14}$O, and $^{14}$N($p,\gamma )^{15}$O reactions rates. In Fig. %
\ref{fig5} the reaction rates of the $^{13}$ N($p,\gamma )^{14}$O, $^{14}$N($%
p,\gamma )^{15}$O and $^{12}$C$(p,\gamma )^{13}$N processes are shown, which
are further used in the calculations of $\tau $. For the $^{13}$N($p,\gamma
)^{14}$O reaction we use results of the present calculations and data from
Ref. \cite{Smith}, for the reaction $^{14}$N($p,\gamma )^{15}$O data \cite%
{29} and \cite{Dubovichenko2020} are used, while for the $^{12}$C$(p,\gamma
)^{13}$N we employed data \cite{29}, which are very close to data given in
the NACRE II database \cite{24}. Let us comment on the difference in the
data for the $^{14}$N($p,\gamma )^{15}$O reaction (curves 3 and 5 in Fig. %
\ref{fig5}). In contrast to Ref. \cite{29}, in Ref. \cite{Dubovichenko2020}
the $^{14}$N($p,\gamma )^{15}$O reaction rate was calculated by taking into
account radiative capture of protons both in the GS of $^{14}$N nucleus and
in all four excited bound levels. Such consideration allows one to describe
experimental data for the astrophysical $S$-factors of the radiative proton
capture on $^{14}$N to five excited states of the $^{15}$O nucleus at the
excitation energies from 5.18 MeV to 6.86 MeV under the assumption, that all
five resonances are $D$ scattering waves. The latter approach leads to a 
significant increase of the $^{14}$N($p,\gamma )^{15}$O reaction rate at
temperatures $T_{9}>0.3$, which is indicated in Fig. \ref{fig5}.

\begin{figure}[t]
\noindent
\begin{centering}
\includegraphics[width=8.8cm]{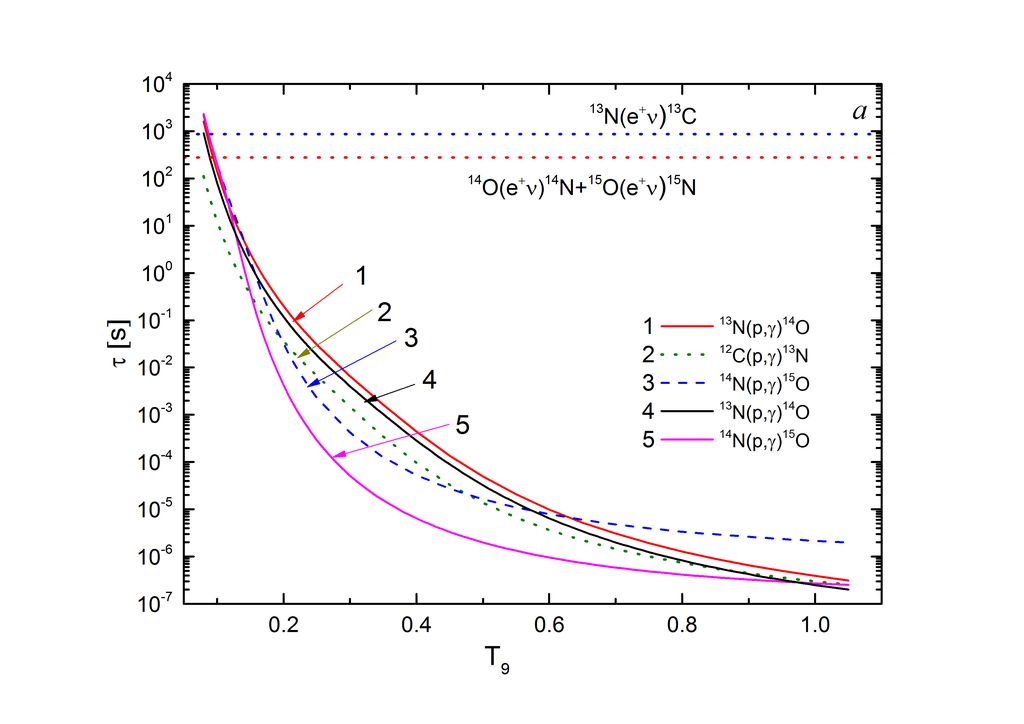}
\includegraphics[width=8.8cm]{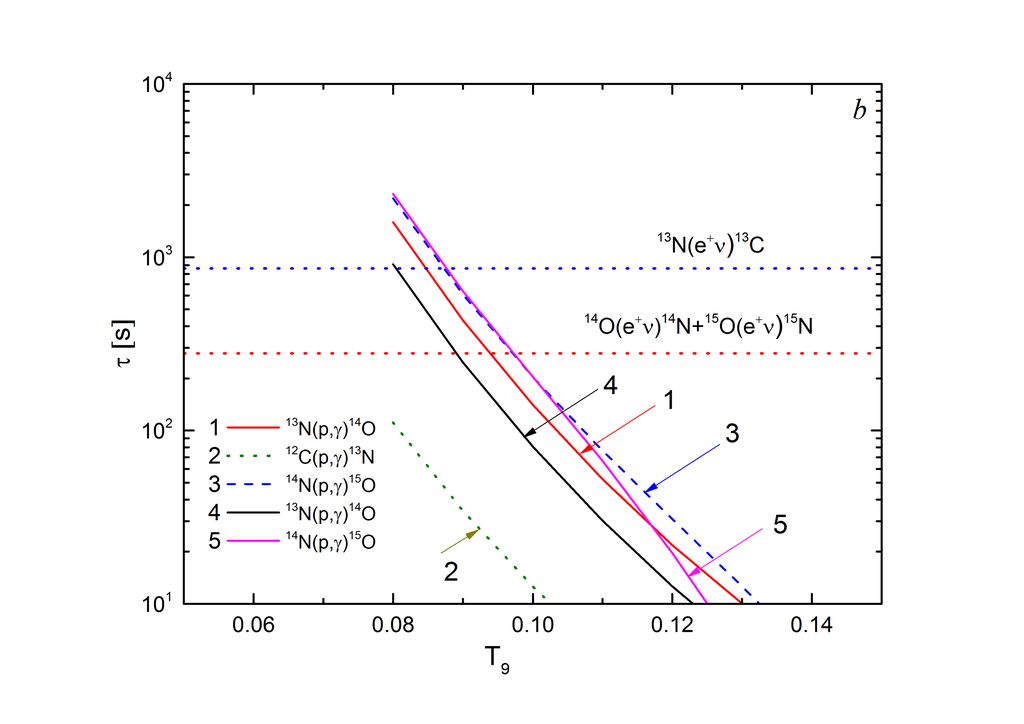}
\par\end{centering}
\caption{(Color online) Comparison of lifetime against hydrogen burning via
the $^{12}$C$(p,\protect\gamma )^{13}$N, $^{13}$N($p,\protect\gamma )^{14}$%
O, and $^{14}$N($p,\protect\gamma )^{15}$O reactions as a function of
temperature, and the $^{13}$N, $^{14}$O, and $^{15}$O $\protect\beta $-
decay lifetimes for the temperature intervals ($a$) $0.08<T_{9}<1.0$ and ($b$%
) $0.08<T_{9}<0.14$. }
\label{fig6}
\end{figure}

\begin{figure}[t]
\noindent
\begin{centering}
\includegraphics[width=10.5cm]{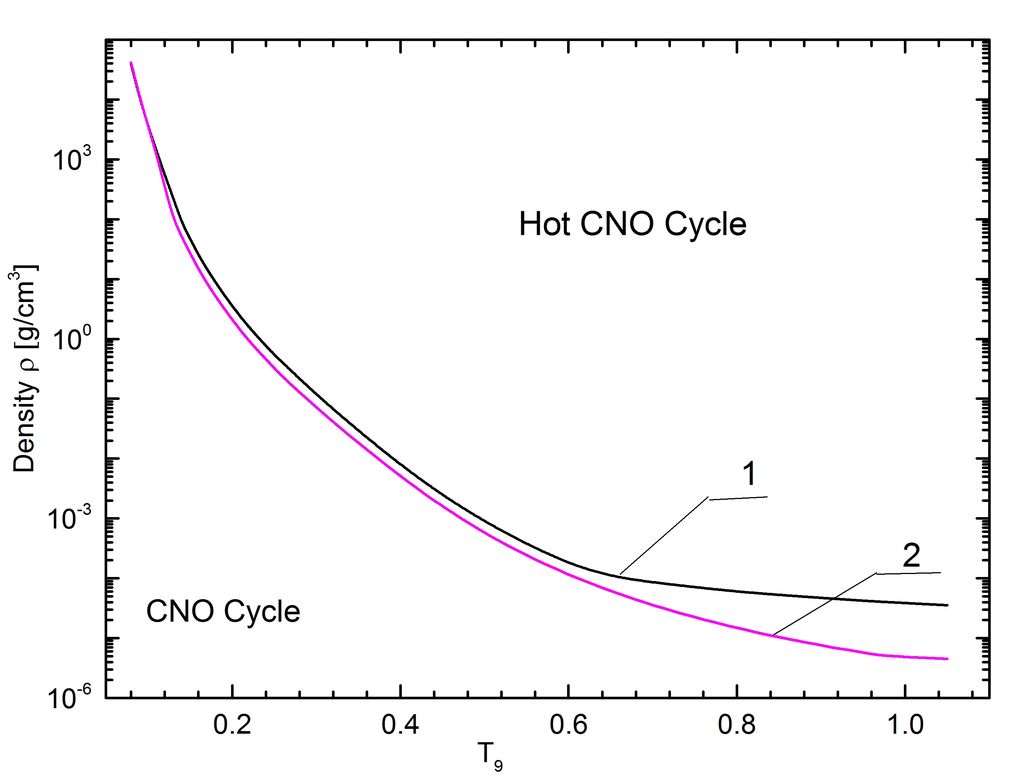}
\par\end{centering}
\caption{(Color online) Density and temperature range for the operation of
the hot CNO cycle. Curves: 1 - result from Ref. \protect\cite{Smith}; 2 -
present result.}
\label{fig7}
\end{figure}

In order to determine the astrophysical temperatures at which the CNO cycle
is converted to the HCNO cycle, it is necessary to determine the $^{13}$N($%
p,\gamma )^{14}$O reaction rate as a function of temperature and compare it
with one for the other processes. Using the reaction rates presented in Fig. %
\ref{fig5}, we calculate the dependence of the lifetime of isotopes
produced in the processes $^{12}$C$(p,\gamma )^{13}$N, $^{13}$N($p,\gamma
)^{14}$O, and $^{14}$N($p,\gamma )^{15}$O on temperature. Following Ref.
\cite{Smith}, in calculations we used for the hydrogen mass fraction $%
X_{H}=0.77$ and the stellar density $\rho =5\times 10^{3}$ g/cm$^{3}$ \cite%
{Wiescher1986}.

The dependencies of the lifetime of isotopes produced in the processes on
the temperature are presented in Fig. \ref{fig6}. The data for the lifetime
of radioactive isotopes are also presented in Fig. \ref{fig6}: $\tau _{^{13}%
\text{N}}$ $=863$ s for the $^{13}$N($e^{+}\nu _{e})^{13}$C, $\tau _{^{14}%
\text{O}}$ $=$ 102 s for the $^{14}$O$(e^{+}\nu _{e})^{14}$N and $\tau
_{^{15}\text{O}}$ $=$176 s for the $^{15}$O($e^{+}\nu _{e})^{15}$N. The
analysis of the results presented in Fig. \ref{fig6} shows that at $T_{9}=$%
0.08 the $^{13}$N($p,\gamma )^{14}$O and $^{13}$N($e^{+}\nu _{e})^{13}$C
reactions have equal lifetime. When the lifetime of $^{14}$O isotope
produced via $^{13}$N($p,\gamma )^{14}$O reaction will be less than the $%
^{13}$N($e^{+}\nu _{e})^{13}$C decay lifetime, the reaction sequence changes
to the hot CNO cycle. For these conditions in CNO cycle the lifetimes of the
$\beta ^{+}$-unstable systems such as $^{13}$N and $^{15}$O are long enough
that proton capture can occur on these unstable nuclei before they undergo
the $\beta ^{+}$-decay.

The onset of the HCNO cycle occurs at $T_{9}=$0.08 when the rate of the
slowest $^{13}$N($p,\gamma )^{14}$O reaction exceeds the $^{14}$O$(e^{+}\nu
_{e})^{14}$N and $^{15}$O($e^{+}\nu _{e})^{15}$N decay rates. Moreover, at $%
T_{9}=0.1$ the ratio of the $^{13}$N($p,\gamma )^{14}$O and $^{13}$N($%
e^{+}\nu _{e})^{13}$C rates is 10.8, in the contract to Ref. \cite{Smith},
where this ratio is about 6. Therefore, at $T_{9}=0.1$ the reaction $^{13}$N(%
$p,\gamma )^{14}$O is already ten times faster than the $^{13}$N($e^{+}\nu
_{e})^{13}$C decay, resulting in the mass flow going via $^{14}$O at the
very onset of the HCNO cycle. The present result indicates that the HCNO
cycle is turned on at the early stage of a nova explosion when the
temperature is lower than reported in the earlier calculations \cite{27} and
\cite{Smith}.

Our calculations lead to the temperature range $0.13<T_{9}<0.97$, where the
reaction rate of $^{14}$N($p,\gamma )^{15}$O is greater than the reaction
rate of $^{13}$N($p,\gamma )^{14}$O. The $^{13}$N($p,\gamma )^{14}$O
reaction rate obtained in the present calculations leads to the temperature
window which is much wider than reported in Ref. \cite{Smith}: $%
0.14<T_{9}<0.64.$ One should mention that the reaction rates for $^{13}$N($%
p,\gamma )^{14}$O in the present work and $^{14}$N($p,\gamma )^{15}$O \cite%
{Dubovichenko2020} are obtained in the framework of the same theoretical
approach.

Following Ref. \cite{Smith}, let's determine the dependence of the stellar
medium density corresponding to the onset of the HCNO cycle on temperature as

\begin{equation}
\rho =\frac{A_{H}}{X_{H}(\tau _{^{14}\text{N}}+\tau _{^{15}\text{N}})}\frac{1%
}{N_{A}\left\langle \sigma _{c}v\right\rangle _{\min }},  \label{Density}
\end{equation}%
where the smallest reaction rate $N_{A}\left\langle \sigma
_{c}v\right\rangle _{\min }$ includes the temperature dependence. An
analysis of the density-temperature relationship allows to determine the
temperatures and densities at which the stellar CNO cycle is converted to
the HCNO cycle. If the density and temperature of the stellar medium fall
above the curve $\rho (T)$ on the density-temperature diagram, then HCNO
cycle occurs, otherwise the CNO cycle operates.

The results of present calculations for the density-temperature dependence $%
\rho (T)$ along with results from Ref. \cite{Smith} are shown in Fig. \ref%
{fig7}. The comparison of our calculations and results \cite{Smith}
indicates that at the same temperature range HCNO cycle operates at the
lower densities of a stellar medium than in the case reported in \cite{Smith}%
. Analysis of the results given at the density-temperature diagram in Fig. %
\ref{fig7} demonstrate that at an early stage of a nova explosion at the
temperature range $0.2$ $T_{9}-0.4$ $T_{9}$ the hot CNO cycle could be
turned on at a twice less density of the stellar matter. The difference
becomes more significant at $T_{9}>0.6$ and the HCNO cycle could be operated
when at $1$ $T_{9}$ a stellar medium density becomes about 10 times less
compared to \cite{Smith}, as can be seen from Fig. \ref{fig7}.

Reanalysis of the astrophysical S-factor and reaction rate of the proton
capture on $^{13}$N nucleus leads us to the numerical differences with
previous studies. These numerical differences bring us to a new temperature
corridor for the conversion of stellar CNO cycle to the HCNO cycle. The
small variation for the range of the HCNO window may lead to the huge
macroscopic consequences on the scale of astrophysical events. Thus, in
supermassive stars at high temperature the ignition of the hot CNO cycle can
occur at much lower densities, generating sufficient energy which can affect
very massive stars collapse at the end of their life cycle.

\section{Conclusion}

We briefly summarize our results. We have employed the modified potential
cluster model to describe the $^{13}$N($p,\gamma $)$^{14}$O reaction at
astrophysical energies and influence of the first $p^{13}$N resonance width
on the astrophysical $S$-factor. At energies of 30--70 keV, the $S$-factor
remains almost constant with the average value 8.4(2) keV$\cdot $b, thereby
determining its value at zero energy, which is determined by the potential
of the $S$-wave scattering. The values of $S(0)$-factor of 7.0(2) to 8.4(2)
keV$\cdot $b are listed in Table \ref{tab1} for three options of potentials,
which correspond to three different values of energies for resonance in the $%
S$ scattering wave. The potentials of the $S$-wave, leading to the correct
resonance width for different resonance energies, do not allow us to obtain
the value of the $S$-factor, which would be consistent with previous
results. Only a decrease in the resonance width to 22--26 keV leads to the $%
S $-factor of the order of 5 keV$\cdot $b, which is consistent with the
upper limit of the results from \cite{24} and the results of other works,
for example, \cite{17,18,25}. Thus, an accurate determination of the width
is crucial.\textbf{\ }Our results demonstrate that contributions of the $M1$
and $E2$ transitions in the $S$-factor are negligible at energies $E<1$ MeV,
but are significant at high energies. At the resonance energy, the $S$%
-factor reaches 2.4 MeV$\cdot $b, which is in a good agreement with the
results of previous studies. Using the MPCM capabilities, it was shown that
the values of the astrophysical $S$-factor of the $^{13}$N($p,\gamma )^{14}$%
O reaction at ultralow energies strongly depends on the $^{3}S_{1}$
resonance parameters.

Based on the potentials for the $S$ scattering wave, consistent with the
energy and widths of the first resonance, the $^{13}$N($p,\gamma $)$^{14}$O
reaction rate was calculated and a simple analytical approximation for the
reaction rate was proposed. The inclusion of resonances at 1.981, 3.117, and
5.123 MeV practically does not affect the reaction rate, although, the
contributions of resonances are clearly visible when calculating the $S$%
-factor. The reason for such a weak influence is their small widths and
relatively large resonance energies. Results of our calculations for the $%
^{13}$N($p,\gamma )^{14}$O reaction rate provide the contribution to the
steadily improving reaction rate libraries.

A precise knowledge of a cross section of the radiative proton capture on $%
^{13}$N isotope at the low energy is important as it plays a key role in the
HCNO cycle, due to the proton capture rate on $^{13}$N at temperature range
of 0.05 $T_{9}-1.0$ $T_{9}$ can become of the same order or larger than the $%
^{13}$N($e^{+}\nu _{e})^{13}$C decay rate. Our calculations show that at $%
T_{9}=0.1$ the ratio of the $^{13}$N($p,\gamma )^{14}$O and $^{13}$N($%
e^{+}\nu _{e})^{13}$C rates is 10.8.

In the context of the CNO cycle scenario, our calculations of the $^{13}$N($%
p,\gamma )^{14}$O and results for the other bottleneck $^{14}$N($p,\gamma
)^{15}$O reaction \cite{Dubovichenko2020} together with the NACRE II data
\cite{24} for the $^{12}$C$(p,\gamma )^{13}$N process show that in the
temperature window $0.13<T_{9}<0.97$, where the reaction rate of $^{14}$N($%
p,\gamma )^{15}$O is greater than the reaction rate of $^{13}$N($p,\gamma
)^{14}$O, occurs the conversion of the CNO cycle to the HCNO cycle. The
present result indicates that the HCNO cycle is turned on at the early stage
of a nova explosion at temperature $T_{9}=$0.08. Therefore, the significant
mass flow through $^{14}$O nucleus begins to occur at temperature $%
T_{9}=0.08 $. Our calculations show that at this temperature the $^{13}$N($%
p,\gamma )^{14}$O reaction rate and the decay rate of the $^{13}$N($e^{+}\nu
_{e})^{13}$C\textit{\ }process are equal.

Our results demonstrate that at early stages of a nova explosion at
temperatures about $0.1$ $T_{9}$ and at late stages of evolution of
supermassive stars at temperatures about $1$ $T_{9}$ the ignition of the hot
CNO cycle could occur at much lower densities of a stellar medium.

Therefore, at temperature and density of a stellar medium such as the
conditions in a nova explosion and very massive stars hydrogen burning
occurs at temperatures $0.01$ $T_{9}-1.0$ $T_{9}.$ For these conditions in
CNO cycle the lifetimes of the $\beta ^{+}$-unstable systems such as $^{13}$%
N and $^{15}$O are long enough that proton capture can occur on these
unstable nuclei before they undergo the $\beta ^{+}$-decay.

\textbf{Acknowledgement}

This work was supported by a grant from the Ministry of Education and
Science of the Republic of Kazakhstan under the program \# BR05236322
\textquotedblleft Investigations of physical processes in extragalactic and
galactic objects and their subsystems\textquotedblright\ under the theme
\textquotedblleft Study of thermonuclear processes in stars and primary
nucleosynthesis of the Universe\textquotedblright\ through the name of
Fesenkov Astrophysical Institute of the National Center for Space Research
and Technology of the Ministry of Digital Development, Innovation and
Aerospace Industry of the Republic of Kazakhstan


\begin{thebibliography}{99}
\bibitem{1} \bigskip E. Wiescher, J. G\"{o}rres, E. Uberseder, G. Imbriani,
and M. Pignatari, Ann. Rev. Nucl. Part. Sci. \textbf{60}, 381 (2010).

\bibitem{Wiescher2012} M. Wiescher, F. K\"{a}ppeler, and K. Langanke, Ann.
Rev. Astron. Astrophys. \textbf{50}, 165 (2012).

\bibitem{Brune2015} C. R. Brune and B. Davids, Ann. Rev. Nucl. Part. Sci.
\textbf{65}, 87 (2015).

\bibitem{Wiescher1980} M. Wiescher, et al., Nucl. Phys. A \textbf{349}, 165
(1980).

\bibitem{2} C. A. Bertulani, A. Gade, Phys. Rep. \textbf{485}, 195 (2010).

\bibitem{Decrock1991} P. Decrock, Th. Delbar, P. Duhamel, W. Galster, M.
Huyse, P. Leleux, I. Licot, et al., Phys. Rev. Lett. \textbf{67}, 808 (1991).

\bibitem{22} P. Decrock, M. Gaelens, M. Huyse, G. Reusen, G. Vancraeynest,
P. Van Duppen, et al., Phys. Rev. C \textbf{48}, 2057 (1993).

\bibitem{Decrock1993} Th. Delbar, W. Galster, P. Leleux, I. Licot, E.
Lienard, P. Lipnik, et al., Phys. Rev. C \textbf{48}, 3088 (1993).

\bibitem{Chunpp1985} T.E. Chupp, R.T. Kouzes, A.B. McDonald, P.D. Parker,
T.F. Wang, A. Howard, Phys Rev. C \textbf{31}, 1023 (1985).

\bibitem{Fernandez1985} P. B. Fernandez, E. G. Adelberger, A. Garcia, Phys.
Rev. C \textbf{40}, 1887 (1989).

\bibitem{Smith} M. S. Smith, P. V. Magnus, K. I. Hahn, R. M. Curley, P. D.
Parker, T. F. Wang, et al., Phys. Rev. C \textbf{47}, 2740 (1993).

\bibitem{Motobayashi1991} T. Motobayashi et al., Phys. Lett. B \textbf{624},
259 (1991).

\bibitem{Kiener1993} J. Kiener et al., Nucl. Phys. A \textbf{552}, 66 \
(1993).

\bibitem{Bauer1994} G. Baur and H. Rebel, J. Phys. G. \textbf{20}, 1 (1994).

\bibitem{16} Z. H. Li et al., Phys. Rev. C \textbf{74}, 035801 (2006).

\bibitem{17} W. Liu et al., Int. J. Mod. Phys. E \textbf{15}, 1899 (2006).

\bibitem{15} R. J. Charity, K. W. Brown, J. Okolowicz, M. Ploszajczak, J. M.
Elson, W. Reviol et al., Phys. Rev. C \textbf{100}, 064305 (2019).

\bibitem{23} P. V. Magnus, E.G. Adelberger, A. Garcia, Phys. Rev. C \textbf{%
49}, R1755 (1994).

\bibitem{27} G. J. Mathews and F. S. Dietrich, Astrophys. J. \textbf{287},
969 (1984).

\bibitem{26} K. Langanke, O. S. Van Rogsmalen and W. A. Fowler, Nucl. Phys.
A \textbf{435}, 657 (1985).

\bibitem{Funck1987} C. Funck and K. Langanke, Nucl. Phys. A \textbf{464}, 90
(1987).

\bibitem{18} X. Tang, A. Azhari, C. Fu, C. A. Gagliardi, A. M.
Mukhamedzhanov, F. Pirlepesov, L. Trache, et al., Phys. Rev. C \textbf{69},
055807 (2004).

\bibitem{181} X. Tang et al., Phys. Rev. C \textbf{67}, 015804 (2003).

\bibitem{25} B. Guo and Z. H. Li, Chin. Phys. Lett. \textbf{24}, 65 (2007).

\bibitem{Huang2010} J. T. Huang, C. A. Bertulani, V. Guimar\~{a}es, Atomic
Data and Nuclear Data Tables \textbf{96}, 824 (2010).

\bibitem{28} C. Angulo et al., Nucl. Phys. A \textbf{656}, 3 (1999).

\bibitem{24} Y. Xu, K. Takahashi, S. Goriely et al., Nucl. Phys. A \textbf{%
918}, 169 (2013).

\bibitem{12} F. Ajzenberg-Selove, Nucl. Phys. A \textbf{523}, 1 (1991).

\bibitem{14} S. I. Sukhoruchkin and Z. N. Soroko, Excited nuclear states,
Sub. G. Suppl. I/25 A-F. Springer, (2016).

\bibitem{3} S. B. Dubovichenko, Thermonuclear processes in Stars and
Universe. Second English ed., expanded and corrected. Germany, Saarbrucken:
Scholar's Press. 2015.

\bibitem{Wiescher2017} M. Wiescher and T. Ahn, Clusters in Astrophysics, in
\textquotedblleft Nuclear Particle Correlations and Cluster
Physics\textquotedblright , Chap. 8, Ed. Wolf-Udo Schr\"{o}der, World
Scientific, pp. 203-255, 2017.

\bibitem{4} S. B. Dubovichenko, Radiative Neutron Capture, Walter de Gruyter
GmbH, Berlin/Boston, 296 p. (2019).

\bibitem{NucPhys2015} S. B. Dubovichenko, A. V. Dzhazairov-Kakhramanov,
Nucl. Phys. A \textbf{941}, 335--363 (2015).

\bibitem{NucPhys2019} S. B. Dubovichenko, N. A. Burkova, A. V.
Dzhazairov-Kakhramanov, R. Ya. Kezerashvili, Ch. T. Omarov, A. S. Tkachenko,
and D. M. Zazulin, Nucl. Phys. A \textbf{987}, 46 (2019).

\bibitem{Dubovichenko2020} S. B. Dubovichenko, N. A. Burkova, and A. V.
Dzhazairov-Kakhramanov, Int. J. Mod. Phys. \textbf{29}, 1930007 (2020).

\bibitem{5} C. A. Barnes, D. D. Clayton, D. N. Schramm, Essays in Nuclear
Astrophysics. Presented to William A. Fowler. UK, Cambridge: Cambridge
University Press. 562p. 1982.

\bibitem{6} V. G. Neudatchin, V. I. Kukulin, V. N. Pomerantsev, and A. A.
Sakharuk, Phys. Rev. C \textbf{45}, 1512 (1992).

\bibitem{7} O. F. Nemets, V. G. Neudatchin, A. T. Rudchik, Yu. F. Smirnov,
Yu. M. Tchuvil'sky, Nucleon association in atomic nuclei and the nuclear
reactions of the many nucleons transfers. Kiev: Naukova Dumka. 488p. 1988.
(in Russian).

\bibitem{Wildermuth} K. Wildermuth and Y. C. Tang, A unified theory of the
nucleus. Braunschweig: Vieweg. 498p., 1977.

\bibitem{RGM} Y. C. Tang, M. LeMere, and D. R. Thompsom, Phys. Rep. \textbf{%
47}, 167 (1978).

\bibitem{8} %
https://physics.nist.gov/cgi-bin/cuu/Value?mudjsearch2520for=atomnuc!

\bibitem{9} http://cdfe.sinp.msu.ru/services/ground/NuclChart\_release.html

\bibitem{Nichitiu1980} F. Nichitiu, Phase shifts analysis in physics.
Romania: Acad. Publ. 416 p. (1980).

\bibitem{10} S. B. Dubovichenko, A. V. Dzhazairov-Kakhramanov, Rus. Phys. J.
\textbf{52}, 833 (2009).

\bibitem{11} V. G. Neudatchin and Yu.F. Smirnov, Nucleon associations in
light nuclei. Moscow: Nauka. 414p. 1969. (in Russian).

\bibitem{111K} V. I. Kukulin, V. G. Neudatchin, Yu. F. Smirnov, Nucl. Phys.
A. \textbf{245}, 429 (1975).

\bibitem{13} C. Itzykson, M. Nauenberg, Rev. Mod. Phys. \textbf{38}, 95
(1966).

\bibitem{Varshalovich} D. A. {Varshalovich, A. N. Moskalev, V. K.
Khersonski, Quantum theory of angular momemtum, World Scientific. 514p.,
1988.}

\bibitem{Tkach2019} A. S. Tkachenko, R. Ya. Kezerashvilic, N. A. Burkova, S.
B. Dubovichenko, Nucl. Phys. A \textbf{991}, 121609 (2019).

\bibitem{Alik46} E. M. Tursunov, S. A. Turakulov, P. Descouvemont, Phys.
Atom. Nucl. \textbf{78}, 193 (2015).

\bibitem{Alik47} F. Hammache et al., Phys. Rev. C \textbf{82}, 065803 (2010).

\bibitem{Fowler} W. A. Fowler, G. R. Caughlan, and B. A. Zimmerman, Ann.
Rev. Astron. Astrophys. \textbf{5,} 525 (1967).

\bibitem{19} A. M. Mukhamedzhanov et al., Nucl. Phys. A \textbf{725}, 279
(2003).

\bibitem{20} G. R. Plattner, R. D. Viollier, Coupling constants of commonly
used nuclear probes, Nucl. Phys. A \textbf{365}, 8 (1981).

\bibitem{Baye2000} D. Baye and E. Brainis, Phys. Rev. C \textbf{61}, 025801
(2000).

\bibitem{Ryan} S. G. Ryan and A. J. Norton, Stellar evolution and
nucleosynthesis, Campridge University Press, New York, 240 p. 2010.

\bibitem{21} http://cdfe.sinp.msu.ru/exfor/index.php.

\bibitem{29} G. R. Caughlan and W. A. Fowler, Atom. Data Nucl. Data Tab.
\textbf{40}, 283 (1988).

\bibitem{Weizsacker} C. F. von Weizs\"{a}cker, Physikalische Zeitschrift.
\textbf{38}, 176 (1937).

\bibitem{Bethe} H. A. Bethe, Phys. Rev. \textbf{55}, 103 (1939).

\bibitem{Rolfs1988} C. Rolfs and W. S. Rodney, Cauldrons in the Cosmos,
University of Chicago Press, Chicago, 1988.

\bibitem{Wiescher1986} M. Wiescher, J. Gorres, F. -K. Thielemann, and H.
Ritter, Astron. Astrophys. \textbf{160}, 56 (1986).
\end{thebibliography}
\end{document}